\documentclass[showpacs,aps,prd,nofootinbib,floatfix,amsmath,amssymb]{revtex4}
\usepackage{graphicx}
\begin{document}

\makeatletter  
\newbox\slashbox \setbox\slashbox=\hbox{$/$}
\newbox\Slashbox \setbox\Slashbox=\hbox{\large$/$}
\def\pFMslash#1{\setbox\@tempboxa=\hbox{$#1$}
  \@tempdima=0.5\wd\slashbox \advance\@tempdima 0.5\wd\@tempboxa
  \copy\slashbox \kern-\@tempdima \box\@tempboxa}
\def\pFMSlash#1{\setbox\@tempboxa=\hbox{$#1$}
  \@tempdima=0.5\wd\Slashbox \advance\@tempdima 0.5\wd\@tempboxa
  \copy\Slashbox \kern-\@tempdima \box\@tempboxa}
\def\FMslash{\protect\pFMslash}
\def\FMSlash{\protect\pFMSlash}
\def\miss#1{\ifmmode{/\mkern-11mu #1}\else{${/\mkern-11mu #1}$}\fi}
\makeatother

\title{Effective Lagrangian description of Higgs mediated flavor violating electromagnetic transitions: implications on lepton flavor violation}
\author{J. I. Aranda$^{(a)}$, A. Flores--Tlalpa$^{(b)}$,  F. Ram\'\i rez--Zavaleta$^{(b)}$, F. J. Tlachino$^{(b)}$, J. J. Toscano$^{(b,c)}$, E. S. Tututi$^{(a)}$}
\address{$^{(a)}$Facultad de Ciencias F\'\i sico Matem\' aticas,
Universidad Michoacana de San Nicol\' as de
Hidalgo, Avenida Francisco J. M\' ujica S/N, 58060, Morelia, Michoac\'an, M\' exico. \\
$^{(b)}$Facultad de Ciencias F\'{\i}sico Matem\'aticas,
Benem\'erita Universidad Aut\'onoma de Puebla, Apartado Postal
1152, Puebla, Puebla, M\'exico.\\
$^{(c)}$Instituto de F\'{\i}sica y Matem\' aticas, Universidad
Michoacana de San Nicol\' as de Hidalgo, Edificio C-3, Ciudad
Universitaria, C.P. 58040, Morelia, Michoac\' an, M\' exico.}
\begin{abstract}
Higgs mediated flavor violating electromagnetic interactions, induced at the one--loop level by a nondiagonal $Hf_if_j$ vertex, with $f_i$ and $f_j$ charged  leptons or quarks, are studied within the context of a completely general effective Yukawa sector that comprises $SU_L(2)\times U_Y(1)$--invariant operators of up to dimension--six. Exact formulae for the one--loop $\gamma f_if_j$ and $\gamma \gamma f_if_j$ couplings are presented and their related processes used to study the phenomena of Higgs mediated lepton flavor violation. The experimental limit on the $\mu \to e\gamma$ decay is used to derive a bound on the branching ratio of the $\mu \to e\gamma \gamma$ transition, which is 6 orders of magnitude stronger than the current experimental limit. Previous results on the $\tau \to \mu \gamma$ and $\tau \to \mu \gamma \gamma$ decays are reproduced. The possibility of detecting signals of lepton flavor violation at $\gamma \gamma$ colliders is explored through the $\gamma \gamma \to l_il_j$ reaction, putting special emphasis on the $\tau \mu$ final state. Using the bound imposed on the $H\tau \mu$ vertex by the current experimental data on the muon anomalous magnetic moment, it is found that about half a hundred events may be produced in the International Linear Collider.
\end{abstract}


\maketitle

\section{Introduction}
\label{in}The International Linear Collider (ILC)~\cite{ILC} is an
ambitious program of electron--positron collisions at the TeV scale,
which will provide a clean environment to make studies beyond the
capabilities of the Large Hadron Collider(LHC). In addition to
complementing and adding precision to the LHC discoveries, the ILC will
provide important information to our knowledge of the standard model
(SM) and promises access to new physics that could eventually show up
at the TeV scale. In particular, the operation of this collider in
the $\gamma \gamma$ mode offers a unique opportunity to explore new
physics effects through production mechanisms that are not
accessible in leptonic or hadronic machines. This collider will
reveal crucial information on those production mechanisms that are
naturally suppressed in electron--positron collisions , as the
involved cross sections can be significantly larger than the
corresponding $e^+e^-$ ones. This is the case of processes that lead to flavor violation, which are quite suppressed within the standard model. In the quark sector of the SM, flavor changing neutral current transitions are very suppressed~\cite{FVSM,Eilam}, whereas the analogous transitions in the lepton sector are absent at any order of perturbation theory. Although absent in the SM, the phenomena of lepton flavor violation (LFV) can arise in many of its  well-motivated extensions. One new and
interesting ingredient of most models beyond the SM is the presence
of more complicated Yukawa sectors, which naturally favor
nondiagonal interactions mediated by the physical Higgs bosons of
the theory. Thus, extended Yukawa sectors are a good place to search
for LFV transitions. In particular, one interesting production
mechanism would be the one leading to the LFV $\tau \mu$ final
state. However, if this transition is mediated by a Higgs boson, one
expects an uninteresting cross section for the $e^+e^-\to \tau \mu$
production mechanism at the tree level and beyond. The reason is
that these types of processes that involve couplings among Higgs
bosons and leptons are naturally scaled by the masses of the
leptons. In this way, the cross section for the $e^+e^-\to \tau \mu$
process is necessarily small, as it is proportional to the product
of the three leptonic masses. Although first generated at the
one--loop level, the $\gamma \gamma \to \tau \mu$ mechanism
offers a more interesting scenario. One advantage consists in the
fact that the cross section depends only of the muon and tau masses,
which are substantially larger than the electron mass. In addition,
as already mentioned, in many cases the $\gamma \gamma$ cross
sections can be much more larger than those associated with $e^+e^-$
collisions.

In this paper we are interested in investigating Higgs mediated flavor violating electromagnetic transitions through the one--loop induced $\gamma f_if_j$ and $\gamma \gamma f_if_j$ couplings and their related processes, namely, the $f_i\to f_j\gamma$ and $f_i\to f_j\gamma \gamma$ decays, as well as the $\gamma \gamma \to f_if_j$ reaction in the context of $\gamma \gamma$ colliders. Although we will present exact expressions for these processes, valid for quarks of leptons, we will focus on their implications on LFV. Although, as already mentioned, these types of processes are absent at any order of perturbation theory within the SM, they can be induced in many of its well-motivated  extensions. In particular, we will exploit the potential of the $\gamma \gamma \to \tau \mu$ reaction at $\gamma \gamma$ colliders to search signals of LFV.  We will focus on
extended Yukawa sectors that are always present within the SM with
additional $SU_L(2)$--Higgs multiplets or in larger gauge groups.
Some processes naturally associated to flavor violation could be significantly impacted by Yukawa sectors
associated with multi--Higgs models, as it is expected that more
complicated Higgs sectors tend to favor this class of new physics
effects. We will assume that the $\gamma f_if_j$ and
$\gamma \gamma f_if_j$ couplings are generated by a virtual scalar field with a
mass of the order of the Fermi scale $v\approx 246$ GeV through the flavor violating $Hf_if_j$ vertex. However,
instead of tackling the problem in a specific model, we will adopt
a model independent approach by using the effective Lagrangian
technique~\cite{EL}, which is an appropriate scheme to study those
processes that are suppressed or forbid in the SM. As it has been
shown in Refs.~\cite{T1,T2}, it is not necessary to
introduce new degrees of freedom in order to generate flavor
violation at the level of classical action; the introduction of
operators of dimension higher than four will be enough. We will
see below that an effective Yukawa sector that incorporates
$SU_L(2)\times U_Y(1)$--invariants of up to dimension-six is
enough to reproduce, in a model-independent manner, the main
features that are common to extended Yukawa sectors, such as the
presence of both flavor violation and CP violation. Although
theories beyond the SM require more complicated Higgs sectors that
include new physical scalars, we stress that our approach for
studying flavor violating electromagnetic couplings induced by a relatively
light scalar particle is sufficiently general to incorporate the
most relevant aspects of extended theories, as in most cases, it
is always possible to identify in an appropriate limit a SM--like
Higgs boson whose couplings to pairs of $W$ and $Z$ bosons
coincide with those given in the minimal SM. Besides its model
independence, our framework has the advantage that it involves an
equal or even smaller number of unknown parameters than those usually
appearing in specific extended Yukawa sectors. As already mentioned, one important goal of this work is to calculate the amplitudes for the $\gamma f_if_j$ and $\gamma \gamma f_if_j$ interactions, which are generated at the one--loop level by the flavor violating $Hf_if_j$ vertex. We will present exact formulae, which means that the masses of all particles will be conserved, no approximations will be made. These general expressions can be used in searching signals of flavor violation both in the quark sector and in the lepton sector. In a previous paper by some of us~\cite{T2}, the implications of Higgs mediated LFV on the tau decays $\tau \to \mu \gamma$ and $\tau \to \mu \gamma \gamma$ were investigated within the context of this 
model-independent description of extended Higgs sectors. In the present paper, we will concentrate in investigating the potential of extended Yukawa sectors to induce lepton flavor violation via the electromagnetic $\mu\to e\gamma$ and $\mu \to e\gamma \gamma$ transitions or through the $\gamma \gamma \to \tau \mu$ reaction, which is the most promising one to detect possible signals of LFV at the ILC.

The paper has been organized as follows. In Sec. \ref{el}, an effective Lagrangian for the Yukawa sector that induces the $Hf_if_j$ vertex is presented. Section \ref{cal} is devoted to calculating the exact amplitudes for the $\gamma f_if_j$ and $\gamma \gamma f_if_j$ couplings. In Sec. \ref{dec}, the branching ratios for the $l_i\to l_j\gamma$ and $l_i\to l_j\gamma \gamma$ decays are studied. In particular, the experimental limit on the $\mu \to e\gamma$ decays will be used to impose a bound on the $\mu \to e\gamma \gamma$ transition. Section \ref{cs} is devoted to calculating and discussing the cross section for the $\gamma \gamma \to \tau \mu$ reaction in the context of the ILC. Finally, in Sec. \ref{co} the conclusions are presented.

\section{Effective Lagrangian description of Higgs mediated flavor violation}
\label{el}In the SM the Yukawa sector is both flavor-- and
CP--conserving, but these effects can be generated at the tree
level if new scalar fields are introduced. One alternative, which
does not contemplate the explicit introduction of new degrees of freedom,
consists in incorporating into the classical action the virtual
effects of the heavy degrees by introducing $SU_L(2)\times
U_Y(1)$--invariant operators of dimension higher than four~\cite{T1,T2}.
Indeed, it is only necessary to extend the Yukawa
sector with dimension--six operators to induce the most general
coupling of the Higgs boson to quarks and leptons. A Yukawa sector
with these features has the following structure~\cite{T1,T2}:
\begin{eqnarray}
{\cal L}^Y_{eff}&=&-Y^l_{ij}(\bar{L}_i\Phi
l_j)-\frac{\alpha^l_{ij}}{\Lambda^2}(\Phi^\dag \Phi)(\bar{L}_i\Phi
l_j)+ H.c. \nonumber \\
&&-Y^d_{ij}(\bar{Q}_i\Phi
d_j)-\frac{\alpha^d_{ij}}{\Lambda^2}(\Phi^\dag \Phi)(\bar{Q}_i\Phi
d_j)+ H.c.\\
&&-Y^u_{ij}(\bar{Q}_i\tilde{\Phi}
u_j)-\frac{\alpha^u_{ij}}{\Lambda^2}(\Phi^\dag
\Phi)(\bar{Q}_i\tilde{\Phi} u_j)+ H.c.\nonumber,
\end{eqnarray}
where $Y_{ij}$, $L_i$, $Q_i$, $\Phi$, $l_i$, $d_i$, and $u_i$
stand for the usual components of the Yukawa matrix, the
left--handed lepton doublet, the left--handed quark doublet, the
Higgs doublet, the right--handed charged lepton singlet, and the
right--handed quark singlets of down and up type, respectively.
The $\alpha_{ij}$ numbers are the components of a $3\times 3$
general matrix, which parametrize the details of the underlying
physics, whereas $\Lambda$ is the typical scale of these new
physics effects.

After spontaneous symmetry breaking, this extended Yukawa sector
can be diagonalized as usual via the unitary matrices
$V^{l,d,u}_L$ and $V^{l,d,u}_R$, which relate gauge states to mass
eigenstates. In the unitary gauge, the diagonalized Lagrangian can
be written as follows:
\begin{eqnarray}
{\cal
L}^Y_{eff}&=&-\Big(1+\frac{g}{2m_W}H\Big)\Big(\bar{E}M_lE+\bar{D}M_dD+\bar{U}M_uU\Big)\nonumber
\\
&&-H\Big(1+\frac{g}{4m_W}H\Big(3+\frac{g}{2m_W}H\Big)\Big)\Big(\bar{E}\Omega^lP_RE+\bar{D}\Omega^d
P_RD+\bar{U}\Omega^u P_RU+H.c.\Big),
\end{eqnarray}
where the $M_a$ ($a=l,d,u$) are the diagonal mass matrix, whereas
$\bar{E}=(\bar{e},\bar{\mu},\bar{\tau})$,
$\bar{D}=(\bar{d},\bar{s},\bar{b})$, and
$\bar{U}=(\bar{u},\bar{c},\bar{t})$ are vectors in the flavor
space. In addition, $\Omega^a$ are matrices defined in the flavor
space through the relation
\begin{equation}
\Omega^a=\frac{1}{\sqrt{2}}\Big(\frac{v}{\Lambda}\Big)^2V^a_L\alpha^aV^{a\dag}_R.
\end{equation}
To generate Higgs--mediated flavor violation at the level of classical
action, it is assumed that neither $Y^{l,d,u}$ nor
$\alpha^{l,d,u}$ are diagonalized by the $V^a_{L,R}$ rotation
matrices, which should only diagonalize the sum
$Y^{l,d,u}+\alpha^{l,d,u}$. As a consequence, mass and
interactions terms would not be simultaneously diagonalized as 
occurs in the dimension--four theory. In addition, if
$\Omega^{a\dag}\neq \Omega^a$, the Higgs boson couples to fermions
through both scalar and pseudoscalar components, which in turn
could lead to CP violation in some processes. As a consequence,
the flavor violating coupling $Hf_if_j$, where $f$ stands for
a charged lepton or quark, has the most general renormalizable
structure of scalar and pseudoscalar type given by
\begin{equation}
-i\Gamma_{ij}=-i(\omega_{ij}P_R+\omega^*_{ij}P_L),
\end{equation}
where $\omega_{i j}=\frac{g\,m_i}{2\,m_W}\delta_{i j}+\Omega_{i j}$, $P_R=\frac{1+\gamma_5}{2}$, and $P_L=\frac{1-\gamma_5}{2}$. This is the vertex that will be used in the next section to calculate the amplitudes for the $\gamma f_if_j$ and $\gamma \gamma f_if_j$ couplings, which are induced at the one--loop level.

To close this section, let us to emphasize that the above
effective Lagrangian describes the most general coupling of
renormalizable type of a scalar field to pairs of fermions, which
reproduces the main features of most of extended Yukawa sectors,
as the most general version of the two--Higgs doublet model
(THDM-III)~\cite{THDM-III} and  multi--Higgs models that comprise
additional multiplets of $SU_L(2)\times U_Y(1)$ or scalar
representations of larger gauge groups. Our approach also covers
more exotic formulations of flavor violation, as the so--called
familon models~\cite{Familons} or theories that involve an
Abelian flavor symmetry~\cite{AFS}. In this way, our results will
be applicable to a wide variety of models that predict
scalar--mediated flavor violation.

\section{The one--loop amplitudes for the $\gamma \gamma f_if_j$ and $\gamma f_if_j$ couplings}
\label{cal} In this section, we present the amplitudes associated with the on--shell $\gamma \gamma f_if_j$ and $\gamma f_if_j$ couplings, which can be used to predict the cross sections or decay widths of their related processes. No approximations will be made and exact formulae will be presented. In particular, the amplitude for the $\gamma \gamma f_if_j$ coupling will be given in terms of scalar products among the four--vectors involved. From these expressions, the conversion to the kinematical variables associated with the scattering $\gamma \gamma \to f_if_j$ process (Mandelstam's variables) or with the three--body $f_i\to f_j\gamma \gamma$ decays (phase space variables) can be easily performed.

\subsection{The $\gamma \gamma f_if_j$ coupling}
\begin{figure}
\centering
\includegraphics[width=3.0in]{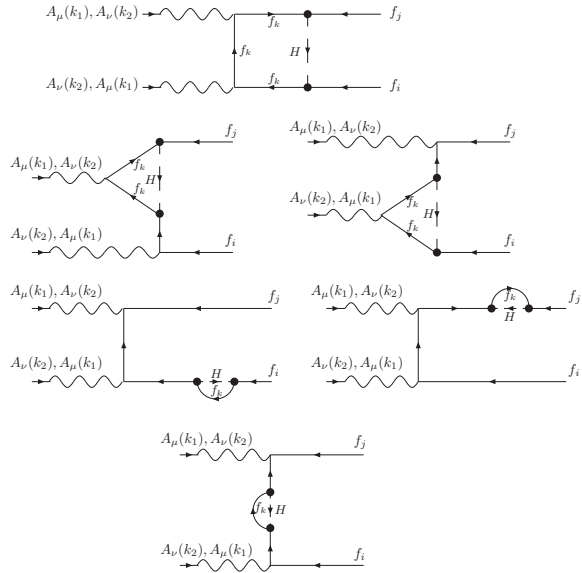}
\caption{\label{BT} Box and reducible diagrams contributing to the $\gamma \gamma f_if_j$ coupling.}
\end{figure}

\begin{figure}
\centering
\includegraphics[width=3.0in]{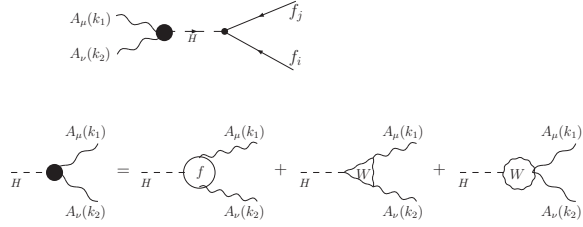}
\caption{\label{HR} Contribution of the standard model loop--induced $H\gamma \gamma$ vertex to the $\gamma \gamma f_if_j$ coupling.}
\end{figure}

The contribution of the flavor violating $Hf_if_j$ vertex to the $\gamma \gamma f_if_j$ coupling is given through the box and reducible diagrams shown in Figs.~\ref{BT} and \ref{HR}, each leading separately to a finite and gauge-invariant result. It should be noticed that the contribution of the $Hf_if_j$ vertex occurs at the one--loop level in graphs of Fig.~\ref{BT}, whereas it contributes at the tree--level in the diagrams of Fig.~\ref{HR}. This last contribution, which is determined by the one--loop $H\gamma \gamma$ vertex, can eventually have a determinant role in the cross section due to an resonant effect of the Higgs boson. In the case of those processes involving light fermions, such as the LFV $\gamma \gamma \to \tau \mu$ reaction, the contributions arising from diagrams of Fig.~\ref{BT} are marginal due to the presence (at first order in the $\Omega_{\tau \mu}$ parameter) of the SM coupling $H\tau \tau$, which is proportional to $m_\tau /v$. However, the contributions given by these diagrams can be competitive with that induced by the diagrams of Fig.~\ref{HR}, as this suppression effect is not present in flavor violating processes involving the quark top~\cite{WP}. We have adopted the convention of taking all momenta incoming: $k_1+k_2+p_i+p_j=0$, where $k_{1,2}$ and $p_{i,j}$ stand for the four--vectors of photons and fermions, respectively. The rest of our notation is indicated in Figs.~\ref{BT} and \ref{HR}.

The analytical structure of the amplitude is dictated by electromagnetic gauge invariance and Bose statistics. In addition, since the structure of the $Hf_if_j$ vertex is of renormalizable type, the amplitude must be free of ultraviolet divergences. We first analyze the contribution given by the set of diagrams shown in Fig.~\ref{BT}. The corresponding amplitude can be written as follows:

\begin{equation}
\mathcal{M}^{\mu\nu}_{\mathrm{Box+red}}=e^2\,\,\overline{\mathrm{v}}_j(p_j)\int\frac{d^{D}k}{(2\pi)^D}
\Bigg[\sum\limits_{n=1}^2\,\frac{\mathcal{T}^{\mu\nu}_{B_n}}{\Delta_{B_n}}
+\sum\limits_{n=1}^4\,\frac{\mathcal{T}^{\mu\nu}_{T_{n}}}{\Delta_{T_n}}
+\sum\limits_{n=1}^6\,\frac{\mathcal{T}^{\mu\nu}_{S_{n}}}{\Delta_{S_n}}\Bigg]\,\mathrm{u}_i(p_i),
\end{equation}
where the subindexes $B$, $T$, and $S$ stand for box, triangle, and self--energy diagrams, respectively. The diverse magnitudes appearing in these expressions are given by
\begin{align}
\mathcal{T}^{\mu\nu}_{B_{1}}&=(\omega_{k j}\,P_R+\omega^*_{k j}\,P_L)\,(\pFMSlash{k}-\pFMSlash{p}_j+m_k)\,\gamma^\nu\,
(\pFMSlash{k}+\pFMSlash{k}_1+\pFMSlash{p}_i+m_k)\,\gamma^\mu\,
(\pFMSlash{k}+\pFMSlash{p}_i+m_k)(\omega_{i k}\,P_R+\omega^*_{i k}\,P_L),\nonumber\\
\mathcal{T}^{\mu\nu}_{B_{2}}&=(\omega_{k j}\,P_R+\omega^*_{k j}\,P_L)\,(\pFMSlash{k}-\pFMSlash{p}_j+m_k)\,\gamma^\mu\,
(\pFMSlash{k}+\pFMSlash{k}_2+\pFMSlash{p}_i+m_k)\,\gamma^\nu\,
(\pFMSlash{k}+\pFMSlash{p}_i+m_k)(\omega_{i k}\,P_R+\omega^*_{i k}\,P_L),\nonumber\\
\mathcal{T}^{\mu\nu}_{T_1}&=(\omega_{k j}\,P_R+\omega^*_{k j}\,P_L)\,(\pFMSlash{k}-\pFMSlash{p}_j+m_k)\,\gamma^\nu\,
(\pFMSlash{k}+\pFMSlash{k}_1+\pFMSlash{p}_i+m_k)\,(\omega_{i k}\,P_R+\omega^*_{i k}\,P_L)\,
(\pFMSlash{k}_1+\pFMSlash{p}_i+m_i)\,\gamma^\mu,\nonumber\\
\mathcal{T}^{\mu\nu}_{T_2}&=(\omega_{k j}\,P_R+\omega^*_{k j}\,P_L)\,(\pFMSlash{k}-\pFMSlash{p}_j+m_k)\,\gamma^\mu\,
(\pFMSlash{k}+\pFMSlash{k}_2+\pFMSlash{p}_i+m_k)\,(\omega_{i k}\,P_R+\omega^*_{i k}\,P_L)\,
(\pFMSlash{k}_2+\pFMSlash{p}_i+m_i)\,\gamma^\nu,\nonumber\\
\mathcal{T}^{\mu\nu}_{T_3}&=\gamma^\nu\,(\pFMSlash{k}_1+\pFMSlash{p}_i+m_j)\,(\omega_{k j}\,P_R+\omega^*_{k j}\,P_L)\,
(\pFMSlash{k}+\pFMSlash{k}_1+\pFMSlash{p}_i+m_k)\,\gamma^\mu\,
(\pFMSlash{k}+\pFMSlash{p}_i+m_k)(\omega_{i k}\,P_R+\omega^*_{i k}\,P_L),\nonumber\\
\mathcal{T}^{\mu\nu}_{T_4}&=\gamma^\mu\,(\pFMSlash{k}_2+\pFMSlash{p}_i+m_j)\,(\omega_{k j}\,P_R+\omega^*_{k j}\,P_L)\,
(\pFMSlash{k}+\pFMSlash{k}_2+\pFMSlash{p}_i+m_k)\,\gamma^\nu\,
(\pFMSlash{k}+\pFMSlash{p}_i+m_k)(\omega_{i k}\,P_R+\omega^*_{i k}\,P_L),\nonumber\\
\mathcal{T}^{\mu\nu}_{S_1}&=\gamma^\mu\,(-\pFMSlash{p}_j-\pFMSlash{k}_1+m_j)\,\gamma^\nu\,
(\pFMSlash{p}_i+m_i)\,(\omega_{k j}\,P_R+\omega^*_{k j}\,P_L)  \,
(\pFMSlash{k}+\pFMSlash{p}_i+m_k)(\omega_{i k}\,P_R+\omega^*_{i k}\,P_L),\nonumber\\
\mathcal{T}^{\mu\nu}_{S_2}&=\gamma^\nu\,(-\pFMSlash{p}_j-\pFMSlash{k}_2+m_j)\,\gamma^\mu\,
(\pFMSlash{p}_i+m_i)\,(\omega_{k j}\,P_R+\omega^*_{k j}\,P_L)  \,
(\pFMSlash{k}+\pFMSlash{p}_i+m_k)(\omega_{i k}\,P_R+\omega^*_{i k}\,P_L),\nonumber\\
\mathcal{T}^{\mu\nu}_{S_3}&=\gamma^\mu\,(-\pFMSlash{p}_j-\pFMSlash{k}_1+m_j)\,(\omega_{k j}\,P_R+\omega^*_{k j}\,P_L)  \,
(\pFMSlash{k}-\pFMSlash{k}_1-\pFMSlash{p}_j+m_k)\,(\omega_{i k}\,P_R+\omega^*_{i k}\,P_L)\,
(-\pFMSlash{k}_1-\pFMSlash{p}_j+m_i)\,\gamma^\nu,\nonumber\\
\mathcal{T}^{\mu\nu}_{S_4}&=\gamma^\nu\,(-\pFMSlash{p}_j-\pFMSlash{k}_2+m_j)\,(\omega_{k j}\,P_R+\omega^*_{k j}\,P_L)  \,
(\pFMSlash{k}-\pFMSlash{k}_2-\pFMSlash{p}_j+m_k)\,(\omega_{i k}\,P_R+\omega^*_{i k}\,P_L)\,
(-\pFMSlash{k}_2-\pFMSlash{p}_j+m_i)\,\gamma^\mu,\nonumber\\
\mathcal{T}^{\mu\nu}_{S_5}&=(\omega_{k j}\,P_R+\omega^*_{k j}\,P_L)  \,(\pFMSlash{k}-\pFMSlash{p}_j+m_k)\,  (\omega_{i k}\,P_R+\omega^*_{i k}\,P_L)\,
(-\pFMSlash{p}_j+m_i)\,\gamma^\nu\,(\pFMSlash{k}_1+\pFMSlash{p}_i+m_i)\,\gamma^\mu,\nonumber\\
\mathcal{T}^{\mu\nu}_{S_6}&=(\omega_{k j}\,P_R+\omega^*_{k j}\,P_L)  \,(\pFMSlash{k}-\pFMSlash{p}_j+m_k)\,  (\omega_{i k}\,P_R+\omega^*_{i k}\,P_L)\,
(-\pFMSlash{p}_j+m_i)\,\gamma^\mu\,(\pFMSlash{k}_2+\pFMSlash{p}_i+m_i)\,\gamma^\nu,
\end{align}

\noindent and

\begin{align}
\Delta_{B_n}&=[(k-p_j)^2-m_k^2][(k+k_n+p_i)^2-m_k^2][(k+p_i)^2-m_k^2][k^2-m_H^2],\nonumber\\
\Delta_{T_n}&=\left\{\begin{array}{ll}
               &[(k-p_j)^2-m_k^2][(k+k_n+p_i)^2-m_k^2][(k_n+p_i)^2-m_i^2][k^2-m_H^2],\hspace{1.5cm} n=1,2 \\
               &[(k_{n-2}+p_i)^2-m_i^2][(k+k_{n-2}+p_i)^2-m_k^2][(k+p_i)^2-m_k^2][k^2-m_H^2],\hspace{0.82cm} n=3,4
             \end{array}\right.,\nonumber\\
\Delta_{S_n}&=\left\{\begin{array}{lll}
               &[(p_j+k_n)^2-m_j^2][p_i^2-m_i^2][(k+p_i)^2-m_k^2][k^2-m_H^2],\hspace{3.3cm} n=1,2 \\
               &[(p_j+k_{n-2})^2-m_j^2][(k-p_j-k_{n-2})^2-m_k^2][(p_j+k_{n-2})^2-m_i^2][k^2-m_H^2],\hspace{0.16cm} n=3,4\\
               &[(k-p_j)^2-m_k^2][p_j^2-m_i^2][(k_{n-4}+p_i)^2-m_i^2][k^2-m_H^2],\hspace{2.95cm} n=5,6
             \end{array}\right..
\end{align}
Gauge invariance means that the amplitude must satisfy the transversality conditions,
\begin{eqnarray}
 &&k _{1\mu}\mathcal{M}^{\mu\nu}_{\mathrm{Box+red}}(k_1,k_2)=0, \\
 &&k_{2\nu}\mathcal{M}^{\mu\nu}_{\mathrm{Box+red}}(k_1,k_2)=0,
\end{eqnarray}
whereas Bose statics requires a symmetric amplitude under the interchanges $k_1\leftrightarrow k_2$ and $\mu \leftrightarrow \nu$:
\begin{equation}
\mathcal{M}^{\mu\nu}_{\mathrm{Box+red}}(k_1,k_2)=\mathcal{M}^{\nu\mu}_{\mathrm{Box+red}}(k_2,k_1).
\end{equation}

To solve the above integrals, we have used the Passarino--Veltman Lorentz tensorial decomposition~\cite{PV} implemented in the FeynCalc computer program~\cite{FC}. Although, in order to make less cumbersome the expressions, we present our results in terms of the form factors that define this Lorentz decomposition, we have verified that they are free of ultraviolet divergences by performing the reduction of these form factors to $B_0$, $C_0$, and $D_0$ scalar functions. Although the box diagrams depend on $B_0$ functions, the divergences cancel among themselves dealing to a finite amplitude. This is in contrast with the sets of triangle and self--energy graphs, which generate divergent pieces that disappear only after adding both sets of amplitudes. This cancellation of ultraviolet divergences is quite intricate due to the fact that the amplitudes depend on the $B_0$ functions in a very complicated way. It is worth mentioning that a similar problem is present in the amplitude for the one--loop $t \to cgg$ decay of the quark top within the context of the SM. The amplitude that presents this problem is induced by the contribution of the pseudo-Goldstone boson associated with the $W$ gauge boson, as it was studied in Ref.~\cite{Eilam}. The authors of this reference verified numerically that their results are free of divergences. In our case, we carried out an analytical analysis in order to convince ourselves that these divergences are absent indeed. To perform this analysis, we replace each of the $B_0$ functions by its divergent part, which is common to all them, namely, we put $B_0\to \Delta=-1/(D-4)$ in our amplitude in order to express the apparently divergent part in a product of the way $FR^{\mu \nu}\Delta$, with $F$ a scalar function that depends in complicated way on kinematics variables and $R^{\mu \nu}$ a nontrivial Lorentz tensor structure. After some nontrivial algebraic manipulations, we achieve the required result $F=0$, which shows that in fact the amplitude is free of ultraviolet divergences. As to gauge invariance, it is achieved only after adding the amplitudes of all diagrams of Fig.~\ref{BT}. On the other hand, the contribution from the set of diagrams shown in Fig.~\ref{HR} is free of ultraviolet divergences and is gauge-invariant, as it is generated essentially by the SM one--loop $H\gamma \gamma$ vertex, which, as  is well--known~\cite{HHG}, is characterized by a finite and gauge-invariant amplitude. After these considerations, the complete amplitude for the $\gamma \gamma f_if_j$ coupling can be conveniently written in the following way:
\begin{align}
\label{am}
\mathcal{M}^{\mu\nu}=\frac{\alpha}{4\,\pi}\,\mathrm{\overline{v}}_j(p_j)\,\Big[
\frac{1}{m_W}\Gamma_{\mathrm{Box}+\mathrm{red}}^{\mu\nu}+\frac{g}{2\,m_W}\Gamma_{ji}\,
\Gamma_{\mathrm{red(H)}}^{\mu\nu}\Big]\,\mathrm{u}_i(p_i)
\end{align}
where $\Gamma_{\mathrm{Box}+\mathrm{red}}^{\mu\nu}$ and $\Gamma_{\mathrm{red(H)}}^{\mu\nu}$ represent the contributions coming from the sets of diagrams given in Figs.~\ref{BT} and \ref{HR}, respectively. We have normalized the amplitudes to the $W$ gauge boson mass $m_W$. The $\Gamma_{\mathrm{Box}+\mathrm{red}}^{\mu\nu}$ tensor amplitude can be organized in terms of 11 gauge structures as follows:
\begin{align}
\Gamma^{\mu\nu}_{\mathrm{Box}+\mathrm{red}}=\sum\limits^{11}_{i=1}\,(T_{i_{R}}^{\mu\nu}\,P_R+T_{i_{L}}^{\mu\nu}\,P_L),
\end{align}
where gauge structures means that they satisfy the transversality conditions
\begin{eqnarray}
&&k_{1\mu} T_{i_{R,L}}^{\mu\nu}=0, \\
&&k_{2\nu} T_{i_{R,L}}^{\mu\nu}=0.
\end{eqnarray}
We have arranged our results in an order that reflects in a manifest way both gauge invariance and Bose symmetry. Accordingly, each of the $T_{i_{R,L}}^{\mu\nu}$ quantities are made of products that contain a gauge structure and a scalar form factor, which respects both symmetries as a whole. These tensor amplitudes are given by:

\begin{align}
T_{1_{L,R}}^{\mu\nu}=&\,F_{1_{L,R}}\,\frac{g^{\mu\nu}\,k_1\cdot k_2-k_1^\nu\,k_2^\mu}{k_1\cdot k_2},\nonumber\\
T_{2_{L,R}}^{\mu\nu}=&\,F_{2_{L,R}}\,\frac{(p_j^\mu\,k_1\cdot k_2-k_2^\mu\,k_1\cdot p_j)(p_j^\nu\,k_1\cdot k_2-k_1^\nu\,k_2\cdot p_j)}{(m_W\,\,k_1\cdot k_2)^2},\nonumber\\
T_{3_{L,R}}^{\mu\nu}=&(F_{3_{L,R}}\,\pFMSlash{k}_1+F_{4_{L,R}}\,\pFMSlash{k}_2)\frac{(g^{\mu\nu}\,k_1\cdot k_2-k_1^\nu\,k_2^\mu)}{m_W\,\,k_1\cdot k_2},\nonumber\\
T_{4_{L,R}}^{\mu\nu}=&\,F_{5_{L,R}}\,\frac{(\gamma^\mu\,k_1\cdot k_2-k_2^\mu\,\pFMSlash{k}_1)(p_j^\nu\,k_1\cdot k_2-k_1^\nu\,k_2\cdot p_j)}{m_W\,\,(k_1\cdot k_2)^2} + \,F_{6_{L,R}}\,\frac{(\gamma^\nu\,k_1\cdot k_2-k_1^\nu\,\pFMSlash{k}_2)(p_j^\mu\,k_1\cdot k_2-k_2^\mu\,k_1\cdot p_j)}{m_W\,(k_1\cdot k_2)^2},\nonumber\\
T_{5_{L,R}}^{\mu\nu}=&(F_{7_{L,R}}\,\pFMSlash{k}_1+F_{8_{L,R}}\,\pFMSlash{k}_2)\frac{(p_j^\mu\,k_1\cdot k_2-k_2^\mu\,k_1\cdot p_j)(p_j^\nu\,k_1\cdot k_2-k_1^\nu\,k_2\cdot p_j)}{m_W^3\,\,(k_1\cdot k_2)^2},\nonumber\\
T_{6_{L,R}}^{\mu\nu}=&\,F_{9_{L,R}}\,\,\frac{\pFMSlash{k}_1\,\gamma^\mu\,(p_j^\nu\,k_1\cdot k_2-k_1^\nu\,k_2\cdot p_j)}{m_W^2\,\,k_1\cdot k_2}+\,F_{10_{L,R}}\,\,\frac{\pFMSlash{k}_2\,\gamma^\nu\,(p_j^\mu\,k_1\cdot k_2-k_2^\mu\,k_1\cdot p_j)}{m_W^2\,\,k_1\cdot k_2},\nonumber\\
T_{7_{L,R}}^{\mu\nu}=&\,F_{11_{L,R}}\,\,\frac{\gamma^\mu\,\pFMSlash{k}_2\,\gamma^\nu\,k_1\cdot k_2-\pFMSlash{k}_1\,\pFMSlash{k}_2\,\gamma^\nu\,k_2^\mu}{m_W\,\,k_1\cdot k_2}+\,F_{12_{L,R}}\,\,\frac{\gamma^\nu\,\pFMSlash{k}_1\,\gamma^\mu\,k_1\cdot k_2-\pFMSlash{k}_2\,\pFMSlash{k}_1\,\gamma^\mu\,k_1^\nu}{m_W\,\,k_1\cdot k_2},\nonumber\\
T_{8_{L,R}}^{\mu\nu}=&\,F_{13_{L,R}}\,\,\frac{\pFMSlash{k}_1\,\gamma^\mu\,\gamma^\nu\,k_1\cdot k_2-\pFMSlash{k}_1\,\gamma^\mu\,\pFMSlash{k}_2\,
k_1^\nu}{m_W\,\,k_1\cdot k_2}+\,F_{14_{L,R}}\,\,\frac{\pFMSlash{k}_2\,\gamma^\nu\,\gamma^\mu\,k_1\cdot k_2-\pFMSlash{k}_2\,\gamma^\nu\,\pFMSlash{k}_1\,
k_2^\mu}{m_W\,\,k_1\cdot k_2},\nonumber\\
T_{9_{L,R}}^{\mu\nu}=&\,F_{15_{L,R}}\,\,\frac{\,\pFMSlash{k}_1\,\pFMSlash{k}_2\,\gamma^\nu\,
(p_j^\mu\,k_1\cdot k_2-k_2^\mu\,k_1\cdot p_j)}{m_W^3\,\,k_1\cdot k_2}+\,F_{16_{L,R}}\,\,\frac{\,\pFMSlash{k}_2\,\pFMSlash{k}_1\,\gamma^\mu\,
(p_j^\nu\,k_1\cdot k_2-k_1^\nu\,k_2\cdot p_j)}{m_W^3\,\,k_1\cdot k_2},\nonumber\\
T_{10_{L,R}}^{\mu\nu}=&\,F_{17_{L,R}}\,\,\frac{\,\pFMSlash{k}_1\,\gamma^\mu\,\pFMSlash{k}_2\,
(p_j^\nu\,k_1\cdot k_2-k_1^\nu\,k_2\cdot p_j)}{m_W^3\,\,k_1\cdot k_2}+\,F_{18_{L,R}}\,\,\frac{\,\pFMSlash{k}_2\,\gamma^\nu\,\pFMSlash{k}_1\,
(p_j^\mu\,k_1\cdot k_2-k_2^\mu\,k_1\cdot p_j)}{m_W^3\,\,k_1\cdot k_2},\nonumber\\
T_{11_{L,R}}^{\mu\nu}=&\,F_{19_{L,R}}\,\,\frac{\pFMSlash{k}_1\,\gamma^\mu\,\pFMSlash{k}_2\,
\gamma^\nu}{m_W^2}+ \,F_{20_{L,R}}\,\,\frac{\pFMSlash{k}_2\,\gamma^\nu\,\pFMSlash{k}_1\,
\gamma^\mu}{m_W^2}.\label{estructuras}
\end{align}
The fact that each $T_{i_{R,L}}^{\mu\nu}$ term respects the Bose symmetry means that, for example, $F_{19_{R,L}}$ transforms into $F_{20_{R,L}}$ under the interchanges $k_1\leftrightarrow k_2$ and $\mu \leftrightarrow \nu$, but $F_{1_{R,L}}$ remains invariant under the same interchanges. On the other hand, the $\Gamma_{\mathrm{red(H)}}^{\mu\nu}$ amplitude is given by

\begin{align}
\Gamma_{\mathrm{red(H)}}^{\mu\nu}=\,&F_{21}\,\frac{k_2^\mu\,k_1^\nu-k_1\cdot
k_2\,g^{\mu\nu}}{2\,k_1\cdot
k_2-m_H^2+i\,m_H\,\Gamma_H}.\label{estructura3}
\end{align}
The form factors $F_{n_{R,L}}$ are presented in the Appendix.

\subsection{The $\gamma f_if_j$ coupling}
\begin{figure}
\centering
\includegraphics[width=3.0in]{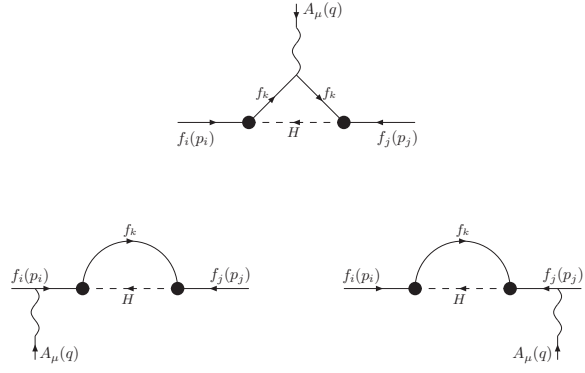}
\caption{\label{TV} Diagrams contributing to the $\gamma f_if_j$ vertex.}
\end{figure}

The contribution of the $Hf_if_j$ vertex to the $\gamma f_if_j$ coupling is given by the diagrams shown in Fig.~\ref{TV}. The amplitude is entirely governed by electromagnetic gauge invariance. In principle, the amplitude could depend on Lorentz structures of the following type: monopole ($\gamma_\mu$), anapole  ($q^2\gamma_\mu-\pFMSlash{q} q_\mu$), magnetic dipole ($\sigma_{\mu \nu}q^\nu$), and electric dipole ($\gamma_5\sigma_{\mu \nu}q^\nu$). However, since this is a flavor violating transition, there is no contribution to the monopole. Also, the anapolar structure is absent in this case because the three particles are on--shell. So, the amplitude must be made of two pieces, one CP--conserving associated with a magnetic dipole of transition and the other being CP--violating, which is associated with an electric dipole of transition. The latter amplitude only can arise if an imaginary part of the $\omega_{ij}$ parameters is assumed. Since the anomalous $Hf_if_j$ vertex has a renormalizable structure, these amplitudes must be free of ultraviolet divergences by renormalization theory. As in the previous case, we will take all momenta incoming, \textit{i.e.}, $p_i+p_j+q=0$, with $q$ the photon four--vector. Then, the vertex associated with the $\gamma f_if_j$ coupling is given by
\begin{equation}
\Gamma_\mu=e\int \frac{d^Dk}{(2\pi)^D}\Bigg(\frac{T^1_\mu}{\Delta_1}+\frac{T^2}{\Delta_2}\frac{-\pFMSlash{p}_j+m_i}{m^2_j-m^2_i}\gamma_\mu
+\gamma_\mu\frac{\pFMSlash{p}_i+m_j}{m^2_i-m^2_j}\frac{T^3}{\Delta_3}\Bigg),
\end{equation}
where
\begin{eqnarray}
T^1_\mu&=&\Gamma_{kj}(\pFMSlash{k}-\pFMSlash{p}_j+m_k)\gamma_\mu (\pFMSlash{k}+\pFMSlash{p}_i+m_k)\Gamma_{ik}, \\
T^2&=&\Gamma_{kj}(\pFMSlash{k}-\pFMSlash{p}_j+m_k)\Gamma_{ik}, \\
T^3&=&\Gamma_{kj}(\pFMSlash{k}+\pFMSlash{p}_i+m_k)\Gamma_{ik},
\end{eqnarray}
\begin{eqnarray}
\Delta_1&=&[k^2-m^2_H][(k+p_i)^2-m^2_k][(k-p_j)^2-m^2_k], \\
\Delta_2&=&[k^2-m^2_H][(k-p_j)^2-m^2_k], \\
\Delta_3&=&[k^2-m^2_H][(k+p_i)^2-m^2_k].
\end{eqnarray}
Once  the integrals are solved, one can see that the dependence on $\gamma_\mu$ cancels exactly after using the Gordon identity, which allows us to eliminate the dependence of the amplitude on the four-vector $p_{i\mu}$. The result can be written in a symmetric form as follows:
\begin{equation}
\label{d1}
\Gamma_\mu=\frac{i\alpha}{8\pi s_W m_W}\Big(\Omega^R_{ij}A(x_i,x_j,x_k)i\sigma_{\mu \nu}q^\nu +
\Omega^I_{ij}\tilde{A}(x_i,x_j,x_k)\gamma_5\sigma_{\mu \nu}q^\nu  \Big),
\end{equation}
where
\begin{eqnarray}
A(x_i,x_j,x_k)&=&(x_i\delta_{ki}+x_j\delta_{kj})F(x_i,x_j,x_k), \\
\tilde{A}(x_i,x_j,x_k)&=&(x_i\delta_{ki}+x_j\delta_{kj})\tilde{F}(x_i,x_j,x_k) ,
\end{eqnarray}
\begin{eqnarray}
\label{d2}
F(x_i,x_j,x_k)&=&\frac{1}{2(x_i+x_j)}\Bigg(-1-x_k(x_i+x_j+x_k)2m^2_HC_0+\frac{2(1-2x_k^2)-2x_k(x_i+x_j)+x_ix_j}{x^2_i-x^2_j}\Big(B_0(1)-B_0(2)\Big)+\nonumber \\
&&\frac{1-x^2_k}{x^2_i-x^2_j}\Bigg(\frac{x_i}{x_j}\Big(B_0(3)-B_0(2)\Big)-\frac{x_j}{x_i}\Big(B_0(3)-B_0(1)\Big)\Bigg)\Bigg),
\end{eqnarray}
\begin{eqnarray}
\label{d3}
\tilde{F}(x_i,x_j,x_k)&=&\frac{1}{2(x_i+x_j)}\Bigg(-1-x_k(x_i-x_j+x_k)2m^2_HC_0+\frac{2(1-2x_k^2)-2x_k(x_i-x_j)-x_ix_j}{x^2_i-x^2_j}\Big(B_0(1)-B_0(2)\Big)+\nonumber \\
&&\frac{1-x^2_k}{x^2_i-x^2_j}\Bigg(\frac{x_i}{x_j}\Big(B_0(2)-B_0(3)\Big)-\frac{x_j}{x_i}\Big(B_0(1)-B_0(3)\Big)\Bigg)\Bigg),
\end{eqnarray}
In the above expressions, $\Omega^R_{ij}=Re(\Omega_{ij})$, $\Omega^I_{ij}=Im(\Omega_{ij})$, and $s_W=\sin\theta_W$. In addition, the dimensionless variables $x_a=m_a/m_H$ were  introduced. Also, $C_0=C_0(m^2_i,m^2_j,0,m^2_k,m^2_H,m^2_k)$, $B_0(3)=B_0(0,m^2_H,m^2_k)$, $B_0(1)=B_0(m^2_i,m^2_H,m^2_k)$, and  $B_0(2)=B_0(m^2_j,m^2_H,m^2_k)$ are Passarino--Veltman scalar functions.

\section{The $l_i\to l_j\gamma$ and $l_i\to l_j\gamma \gamma$ decays}
\label{dec} In this section, we analyze the branching ratios for the $l_i\to l_j\gamma$ and $l_i\to l_j\gamma \gamma$ decays and confront our prediction with  current experimental limits. We will concentrate in the  $\mu\to e\gamma $ and $\mu \to e\gamma \gamma$ transitions, as those involving the $\tau$ lepton were already studied in Ref.~\cite{T2}, in which the limits $Br(\tau \to \mu \gamma)<7.5 \times 10^{-10}$ and $Br(\tau \to \mu \gamma \gamma)<2.3\times 10^{-12}$ were derived.

\subsection{Experimental constraints from $l_i\to l_j\gamma$ and $\mu \to e\bar{e}e$}
From Eqs.~(\ref{d1}, \ref{d2}, \ref{d3}), it is straightforward to construct the branching ratio for the $f_i\to f_j\gamma$ decay. It can be written as:
\begin{equation}
\label{br2b}
Br(f_i\to f_j\gamma)=\frac{\alpha^2}{2^9\pi^3s^2_W}\Big(\frac{m_i}{\Gamma_i}\Big)\Big(\frac{m_H}{m_W}\Big)^2x^2_i\Bigg(1-\frac{x^2_j}{x^2_i}\Bigg)^3
\Bigg((\Omega^R_{ij})^2|A(x_i,x_j)|^2+(\Omega^I_{ij})^2|\tilde{A}(x_i,x_j)|^2\Bigg),
\end{equation}
where $\Gamma_i$ is the total decay width of the fermion $f_i$. This is an exact result, which can be used for studying Higgs mediated flavor violation in both the quark and the lepton sectors, but in this paper we will focus on LFV transitions. In particular, it is interesting to use the current experimental limits on the LFV $l_i\to l_j\gamma$ transitions to constrain the $\Omega_{ij}$ parameters. We will focus on the LFV transitions of the muon. It is important to comment at this point that the complex component of $\Omega_{\mu e}$ also contributes to the electric dipole moment ($d_e$) of the electron. It can be shown that the current limit on $d_e$ induces a very strong bound on $\Omega^I_{\mu e}$~\cite{PT}, so from now on this contribution will be neglected.

The limit on the branching ratio of the $\mu \to e\gamma$ decay reported by the Particle Data Group~\cite{PDG} is
\begin{equation}
Br_{exp}(\mu \to e\gamma)<1.2\times 10^{-11}.
\end{equation}
Then, working out at first order in the $\Omega^R_{\mu e}\equiv \Omega_{\mu e}$ parameter, one obtains the following constraint
\begin{equation}
\label{b1}
\Omega_{\mu e}<7\times 10^{-3}.
\end{equation}
This bound was obtained assuming a value for the Higgs mass of about  the lower limit imposed by LEP on the Higgs mass~\cite{LEPBound}. In the next subsection, we will use this constraint to impose a bound on the $\mu \to e\gamma \gamma$ decay.

\begin{figure}
\centering
\includegraphics[width=3.0in]{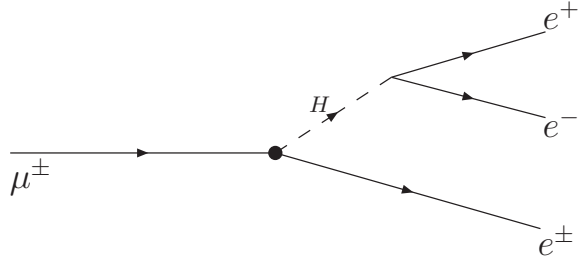}
\caption{\label{HR1} Diagram contributing to the $\mu \to e\bar{e}e$ decay.}
\end{figure}

\begin{figure}
\centering
\includegraphics[width=3.0in]{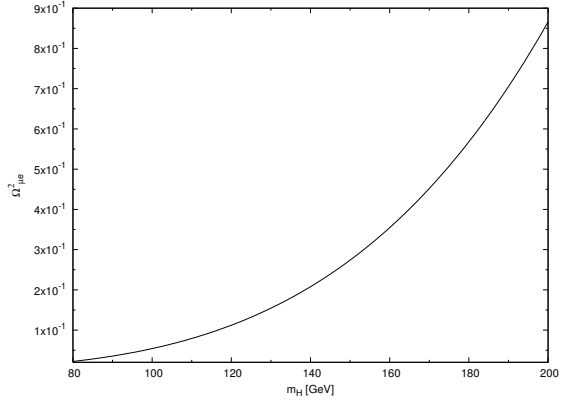}
\caption{\label{H} Behavior of $|\Omega_{\mu e}|^2$ as a function of the Higgs mass.}
\end{figure}

It is interesting to compare this bound with the one that can be derived from the experimental limit on the branching ratio of the tree--body decay $\mu \to eee$, which is strongly restricted by the experiment. In the context of our effective theory, this decay can occur through a virtual Higgs boson as it is shown in Fig.~\ref{HR1}. The amplitude is governed by the LFV $H\mu e$ vertex and the SM one $Hee$, which is very suppressed. The corresponding branching ratio can be written as follows:
\begin{equation}
Br(\mu \to eee)=\frac{\alpha |\Omega{\mu e}|^2}{256\pi^2s^2_W}\Big(\frac{m_e}{m_W}\Big)^2\Big(\frac{m_\mu}{\Gamma_\mu}\Big)f(y,z),
\end{equation}
where
\begin{equation}
f(y,z)=\int^1_0dx\frac{z(1-x)^2}{(x-y)^2+yz}.
\end{equation}
In the above expressions, $y=m^2_H/m^2_\mu$ and $z=\Gamma^2_H/m^2_\mu$, where $\Gamma_\mu$ and $\Gamma_H$ are the muon and Higgs total decay widths, respectively. To bound the $\Omega_{\mu e}$ parameter, we use the limit for $Br(\mu \to eee)$ reported by the Particle Data Group~\cite{PDG}, which is given by
\begin{equation}
Br_{Exp}(\mu \to eee)<10^{-12}.
\end{equation}
In Fig.~\ref{H} the behavior of $|\Omega_{\mu e}|^2$ as a function the  Higgs mass is displayed. It can be appreciated from this figure that $|\Omega_{\mu e}|<0.3$, at the best. This bound is almost 2 orders of magnitude less stringent than that derived above from the experimental limit on the $\mu \to e\gamma$ decay. The important difference between these bounds can be understood by noting that while the $\mu \to eee$ decay is governed by the LFV $H\mu e$ vertex and the SM one $Hee$, the $\mu \to e\gamma$ transition is determined by the LFV $H\mu e$ vertex and the SM one $H\mu \mu$. Besides a phase space factor, the branching ratio for the $\mu \to eee$ decay is suppressed by an additional $(m_e/m_\mu)^2\approx 2.5\times 10^{-5}$ factor with respect to the one associated to the two--body $\mu \to e\gamma$ decay.

\subsection{The $l_i\to l_j\gamma \gamma$ decay}
 The kinematical of the $l_i\to l_j\gamma \gamma$ decay is characterized by three dimensionless variables $x$, $y$, and $z$ ($x+y+z=2$), which are related to the involved four--vector scalar products as follows:
\begin{align}\label{escalaresdec}
k_1 \cdot p_i &= -\frac{x\,m_i^2}{2},\nonumber\\
k_2 \cdot p_i &= -\frac{y\,m_i^2}{2},\nonumber\\
p_i \cdot p_j &= -\frac{z\,m_i^2}{2},\nonumber\\
k_1 \cdot k_2 &= \frac{(x + y + \xi - 1)\,m_i^2}{2},\nonumber\\
k_1 \cdot p_j &= \frac{(1 - \xi - y)\,m_i^2}{2},\nonumber\\
k_2 \cdot p_j &= \frac{(1 - \xi - x)\,m_i^2}{2},
\end{align}
where $\xi = m_j^2/{m_i^2}$.

\begin{figure}
\centering
\includegraphics[width=3.0in]{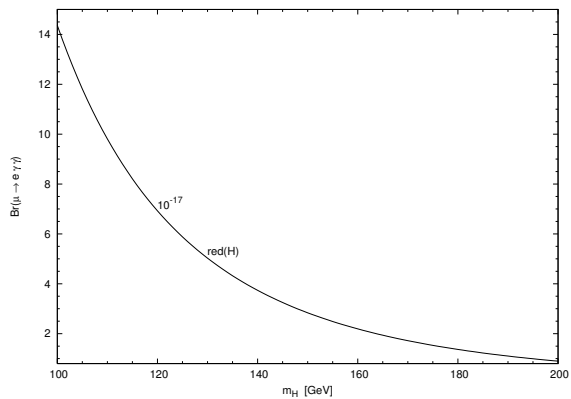}
\caption{\label{B} Behavior of $Br(\mu \to e\gamma \gamma)$ as a function of the Higgs mass.}
\end{figure}

The decay width for the three--body $l_i\to l_j\gamma \gamma$ transition is determined by the squared of the amplitude given in Eq.~(\ref{am}). The corresponding branching ratio is given by
\begin{align}
\label{decay_width} \mathrm{Br}(l_i\to l_j\gamma\gamma)=\frac{1}{256\pi^3}\,\frac{m_i}{\Gamma_i}\,\int\limits_{2\sqrt{\xi}}^{1+\xi}
dx\int\limits_{y_{-}}^{y_+}dy\,|{\cal {M}}|^2,
\end{align}
where ${\cal M}={\cal M}^{\mu \nu}\epsilon^*_\mu( k_1,\lambda_1)\epsilon^*_\nu (k_2,\lambda_2)$, with $\epsilon^*_\mu( k_1,\lambda_1)$ and $\epsilon^*_\nu (k_2,\lambda_2)$ the polarization vectors of the photons. The integration limits are given by
\begin{eqnarray}
y_+&=&\frac{1}{2}\left[(2-x)+\sqrt{x^2-4\,\xi}\right], \\
y_-&=&\frac{1}{2}\left[(2-x)-\sqrt{x^2-4\,\xi}\right].
\end{eqnarray}
The $\tau \to \mu \gamma \gamma$ decay has already been studied in Ref.~\cite{T2}. Here, we are interested in obtaining a bound on the $\mu \to e\gamma \gamma$ transition using the limit on the $\Omega_{\mu e}$ parameter obtained above. The behavior of the corresponding branching ratio is shown in Fig.~\ref{B} as a function of the Higgs mass. It can be appreciated from this figure that this branching ratio ranges from approximately $10^{-16}$ to $10^{-17}$. Conservatively, we can assume the following bound:
\begin{equation}
Br(\mu \to e\gamma \gamma)<10^{-16},
\end{equation}
which would be the highest branching ratio allowed by the lower bound imposed by LEP~\cite{LEPBound} on the Higgs mass. This means that the $\mu \to e\gamma \gamma$ transition may be undetectable if it is mediated by a Higgs boson. This bound for $Br(\mu \to e\gamma \gamma)$ is almost 6 orders of magnitude more stringent than the experimental limit reported by the Particle Data Group~\cite{PDG}, which is $7.2\times 10^{-11}$. In obtaining the above constraint, only the contribution of diagrams given in Fig.~\ref{HR} were considered, as the contribution of the set of diagrams shown in Fig.~\ref{BT} is insignificant in this process involving very light fermions.

\section{The cross section for the $\gamma \gamma \to \tau \mu$ process}
\label{cs}In this section, we analyze the possibility of detecting signals of LFV at the ILC through the $\gamma \gamma \to \tau \mu$ reaction. The constraint on the $H\mu e$ coupling derived in the previous section from the experimental limit on the $\mu \to e\gamma$ decay is so restrictive that no signals can be detected via the $\gamma \gamma \to  \mu e$ process. On the other hand, to predict the cross section for the $\gamma \gamma \to \tau \mu$ reaction, we will need to assume some value for the $\Omega_{\tau \mu}$ parameter. As in the case of the muon decays studied above, we will assume that the imaginary part of $\Omega_{\tau \mu}$ is very suppressed compared with its real part, so from now on $\Omega_{\tau \mu}$ stands for the real part of this parameter. There are various experimental limits~\cite{PDG} which can be used to bound $\Omega_{\tau \mu}$, but the best constraint arises from the anomalous magnetic moment of the muon, which leads to the following bound~\cite{PT}:
\begin{equation}
\Omega^2_{\tau \mu}<1.1\times 10^{-3},
\end{equation}
for $115$ GeV $<m_H<200$ GeV.

The kinematic of the $\gamma \gamma \to \tau \mu$ process is determined by  Mandelstam's variables, which are given by

\begin{align}\label{escalaresdis}
k_1\cdot k_2&=\hat s,\nonumber\\
k_1\cdot p_\mu&=\frac{\hat t-m_\mu^2}{2},\nonumber\\
k_2\cdot p_\mu&=\frac{\hat u-m_\mu^2}{2},\nonumber\\
k_1\cdot p_\tau&=\frac{\hat u-m_\tau^2}{2},\nonumber\\
k_2\cdot p_\tau&=\frac{\hat t-m_\tau^2}{2}.
\end{align}
The unpolarized cross section for this process can be written as
\begin{equation}
\hat{\sigma}(\gamma \gamma \to \tau \mu)=\frac{1}{16\,\pi^2\,
\hat{s}^2}\int^{\hat{t}_{max}}_{\hat{t}_{min}}d\hat{t}\, |{{\cal M}}|^2,\label{t}
\end{equation}
 where ${\cal M}$ is the invariant amplitude which is given by Eq.~(\ref{am}). In the above expression, the integration limits are given by
\begin{equation}
\hat{t}_{max}(\hat{t}_{min})=-\frac{\hat s}{2}\left(1-\frac{m_\mu^2+m_\tau^2}{\hat s}\right)\left(1\mp\sqrt{1-\delta}\right),\label{tmax}
\end{equation}
where
\begin{equation}
\delta=\left(\frac{2\,m_\mu\,m_\tau}{\hat s-m_\mu^2-m_\tau^2}\right)^2.
\end{equation}
On the other hand, the convoluted cross section, \textit{i.e.}, the cross section for the complete process $e^+e^-\to \gamma \gamma \to \tau \mu$, can be written as
\begin{equation}
\sigma(s)=\int^{y_{max}}_{(m_\mu+m_\tau)/\sqrt{s}}dz\,\frac{d{\cal
L}_{\gamma \gamma}}{dz}\,\hat{\sigma}(\gamma \gamma \to \tau \mu),
\end{equation}
where $\hat{s}=z^2s$, with $\sqrt{s}(\sqrt{\hat{s}})$ the center of mass energy of the $e^+e^-(\gamma \gamma)$ collision, and $\frac{d{\cal L}_{\gamma \gamma}}{dz}$ is the luminosity of the photons, defined as
\begin{equation}
\label{l1} \frac{d{\cal L}_{\gamma
\gamma}}{dz}=2z\int^{y_{max}}_{z^2/y_{max}}\frac{dy}{y}f_{\gamma/e}(y)f_{\gamma/e}(z^2/y),
\end{equation}
where the energy spectrum for the backscattered photon is given by~\cite{TEL}
\begin{equation}
\label{l2}
f_{\gamma/e}(y)=\frac{1}{D(\chi)}\Big[1-y+\frac{1}{1-y}-\frac{4y}{\chi
(1-y)}+\frac{4y^2}{\chi^2(1-y)^2}\Big],
\end{equation}
with
\begin{equation}
\label{l3}
D(\chi)=\Big(1-\frac{4}{\chi}-\frac{8}{\chi^2}\Big)\log(1+\chi)+\frac{1}{2}+\frac{8}{\chi}-\frac{1}{2(1+\chi)^2}.
\end{equation}
In this expression, $\chi=(4E_0\omega_0)/m^2_e$, where $m_e$ and $E_0$ are the mass and energy of the electron, respectively; $\omega_0$ is the laser--photon energy; and $y$ represents the fraction of the energy of the incident electron carried by the backscattered photon. The optimum values for the $y_{max}$ and $\chi$ parameters are  $y_{max}\approx 0_\cdot83$ and $\chi=2(1+\sqrt{2})$.

\subsection{Discussion}
We now turn to discuss our results within the experimental context of the ILC. Although this collider is intended to operate initially at a center of mass energy of $500$ GeV with a luminosity of $\mathcal{L}_{e^+e^-}=500$ $fb^{-1}$~\cite{ILC}, it is contemplated to increase the energy up to about $1$ TeV in a subsequent stage. In the following, we will to carry out our numerical analysis in the interval $500$ GeV$<\sqrt{s}<3000$ GeV. We will discuss separately the contributions to this process coming from diagrams displayed in Fig.~\ref{BT} and Fig.~\ref{HR}. To analyze numerically our results, we will work at first order in the anomalous parameter $\Omega_{\tau \mu}$. We will use the value determined by the upper limit imposed by the anomalous magnetic moment of the muon, namely, $\Omega^2_{\tau \mu}<1.1\times 10^{-3}$~\cite{PT}. It turns out to be that those contributions coming from diagrams shown in Fig.~\ref{BT} are proportional to a sum of two amplitudes, each characterized by the factors $(m_\tau/v)(\Omega_{\tau \mu})$ and $(m_\mu/v)(\Omega_{\tau \mu})$, which determine the intensity of the SM coupling $Hl_il_i$ and the LFV one $Hl_il_j$. Such amplitudes represent the two possible situations that can occur when  the $\tau$ or $\mu$ particles circulate in 
the loops. Both  such possibilities are implicit in our general and exact calculation represented by the amplitude given by Eq.~(\ref{am}), but we will neglect the contribution associated with a virtual muon, as it is quite suppressed with respect to the corresponding tau contribution.

\begin{figure}
\centering
\includegraphics[width=3.5in]{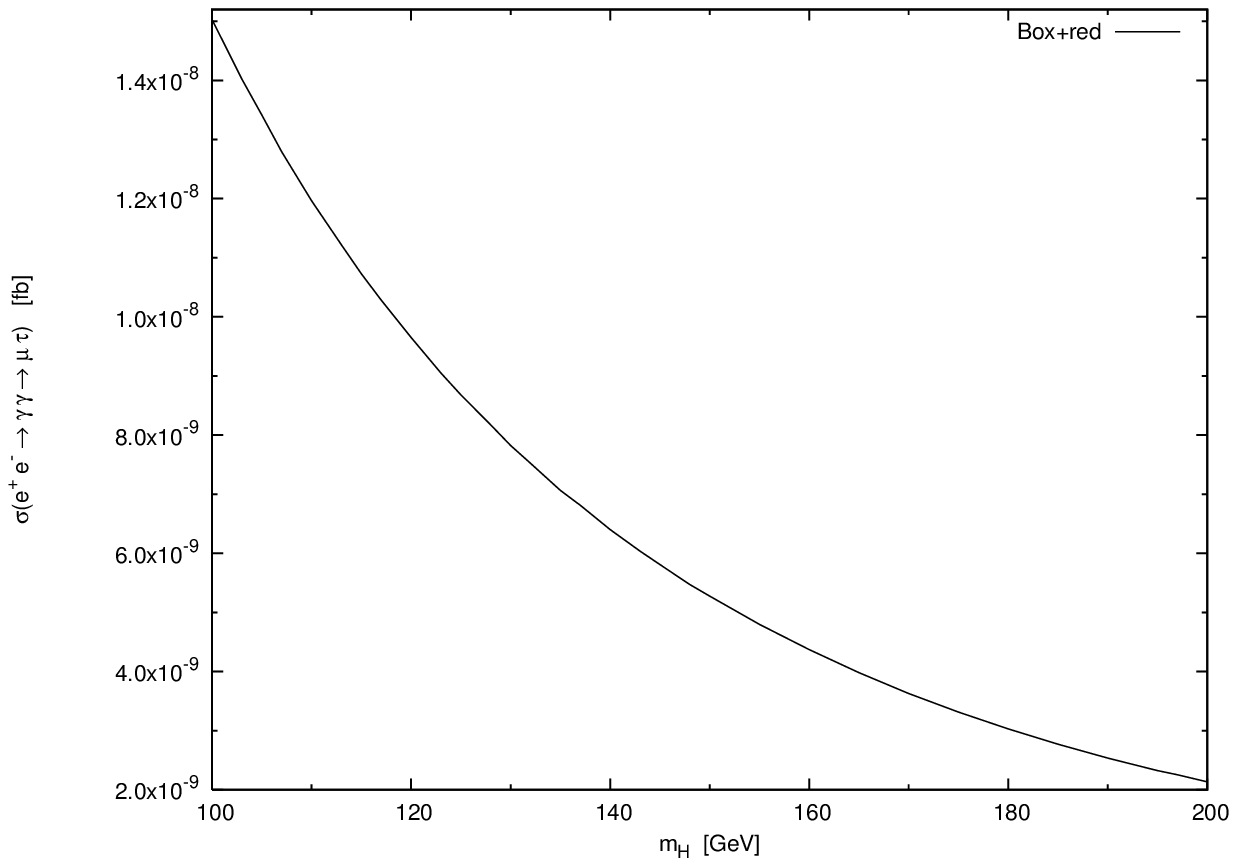}
\includegraphics[width=3.5in]{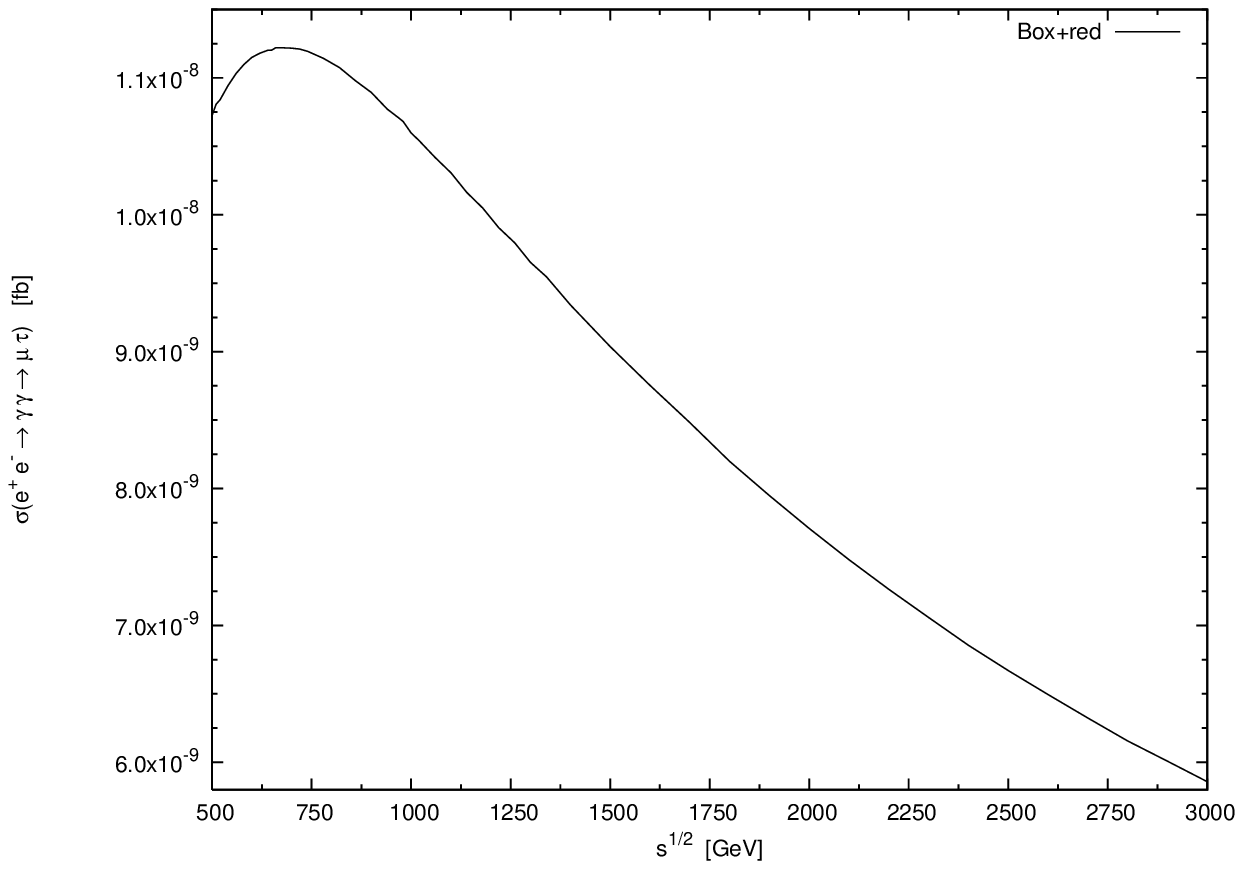}
\caption{\label{BTD}Contribution of diagrams of Fig.~\ref{BT} to the cross section $\sigma(e^+e^-\to\gamma\gamma\to \mu\tau)$ as a function of the Higgs mass for $\sqrt{s}=500$ GeV (left) and as a function of $\sqrt{s}$ for $m_H=115$ GeV (right).}
\end{figure}

\begin{figure}
\centering
\includegraphics[width=3.5in]{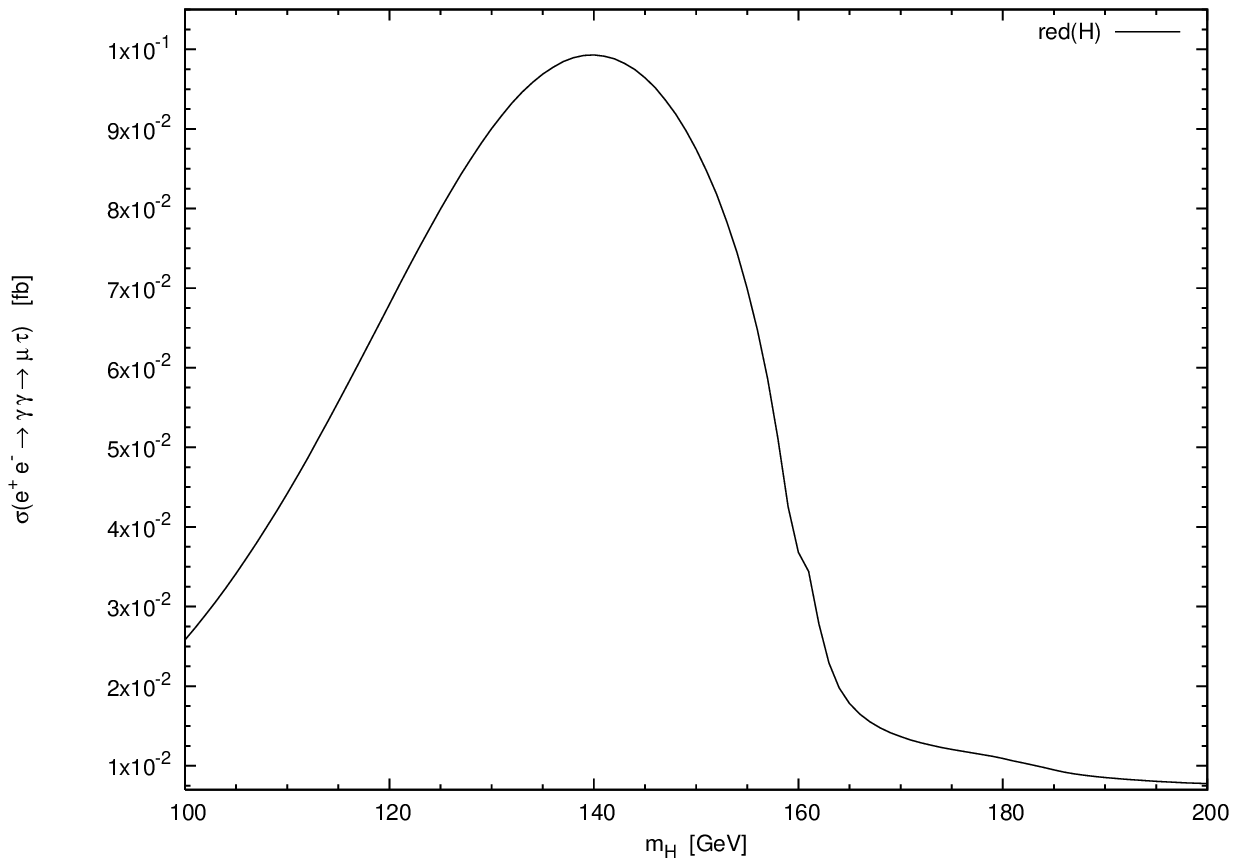}
\includegraphics[width=3.5in]{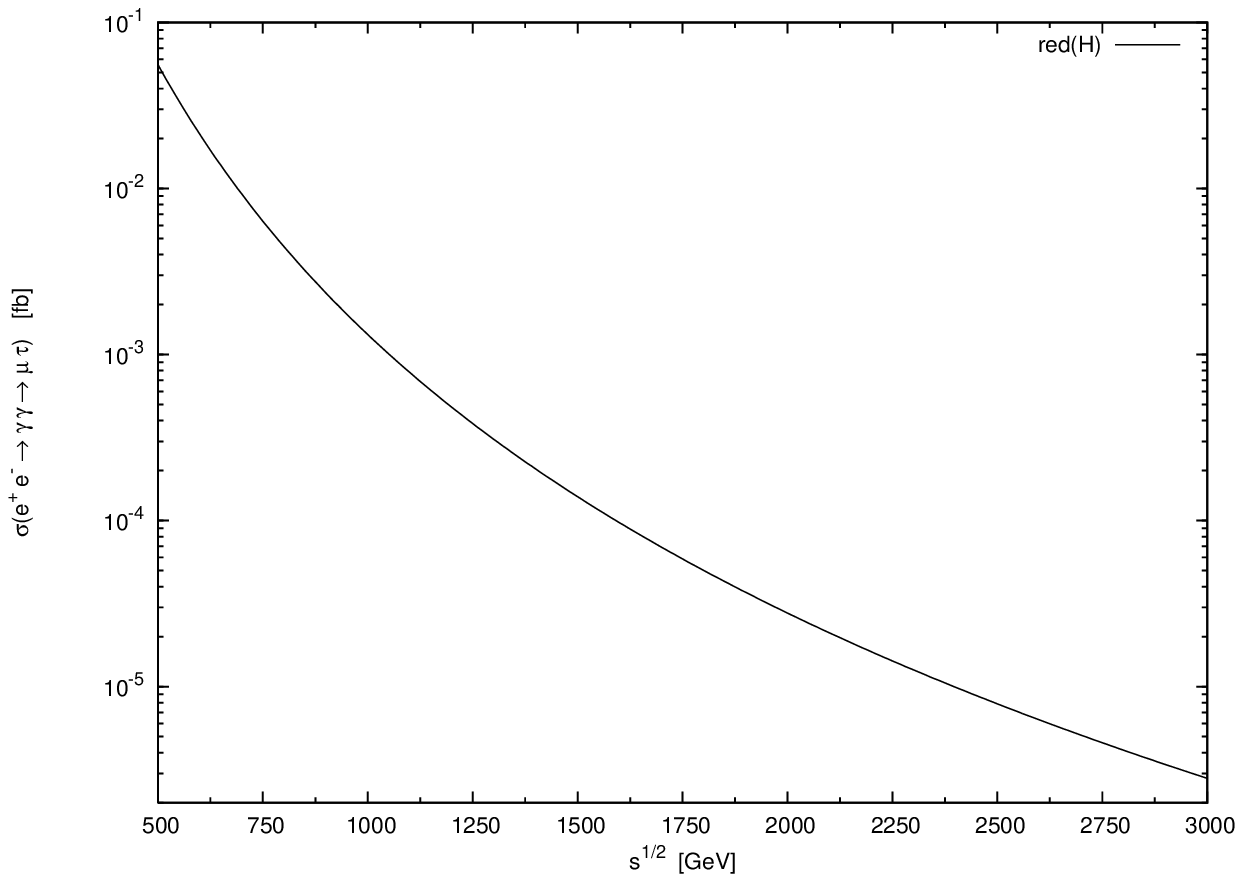}
\caption{\label{HRD}Contribution of diagrams of Fig.~\ref{HR} to the cross section $\sigma(e^+e^-\to\gamma\gamma\to \mu\tau)$ as a function of the Higgs mass for $\sqrt{s}=500$ GeV (left) and as a function of $\sqrt{s}$ for $m_H=115$ GeV (right).}
\end{figure}

\begin{figure}
\centering
\includegraphics[width=3.5in]{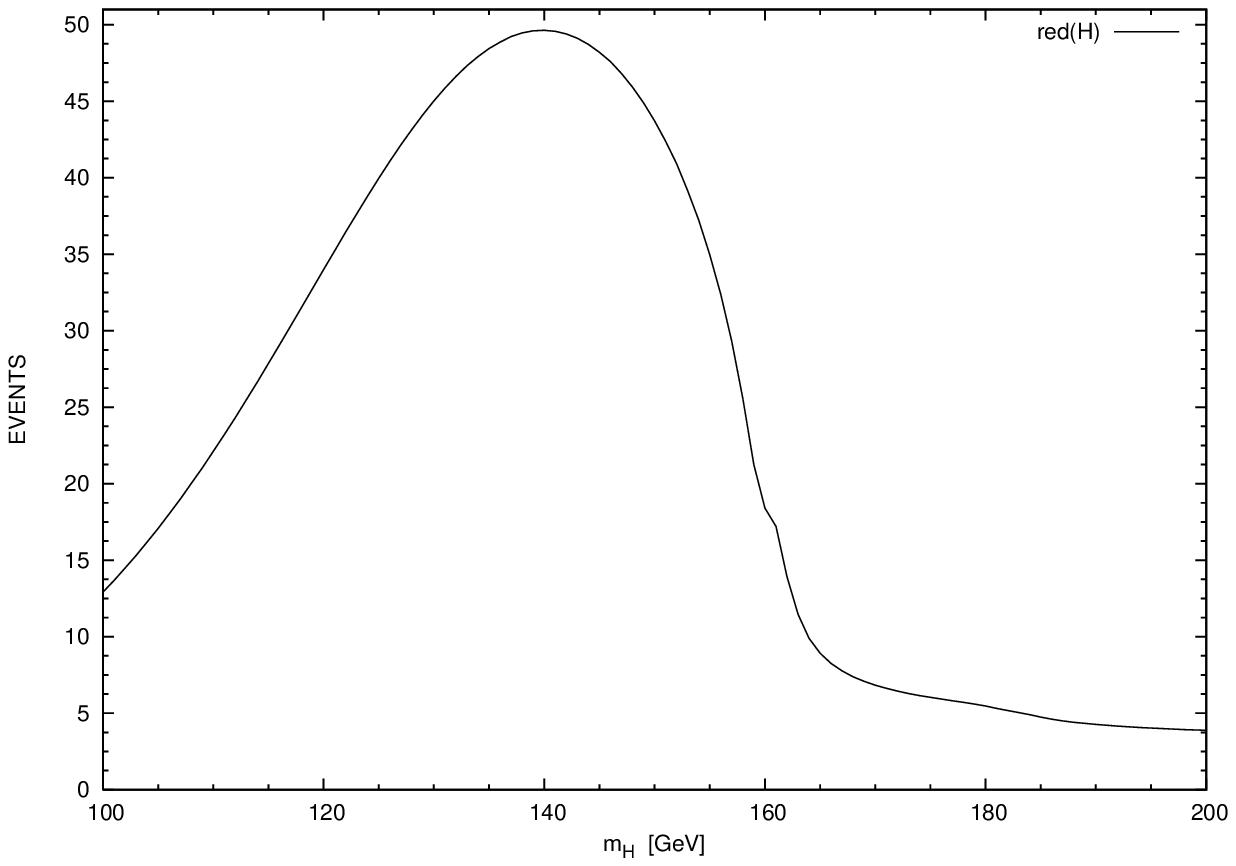}
\includegraphics[width=3.5in]{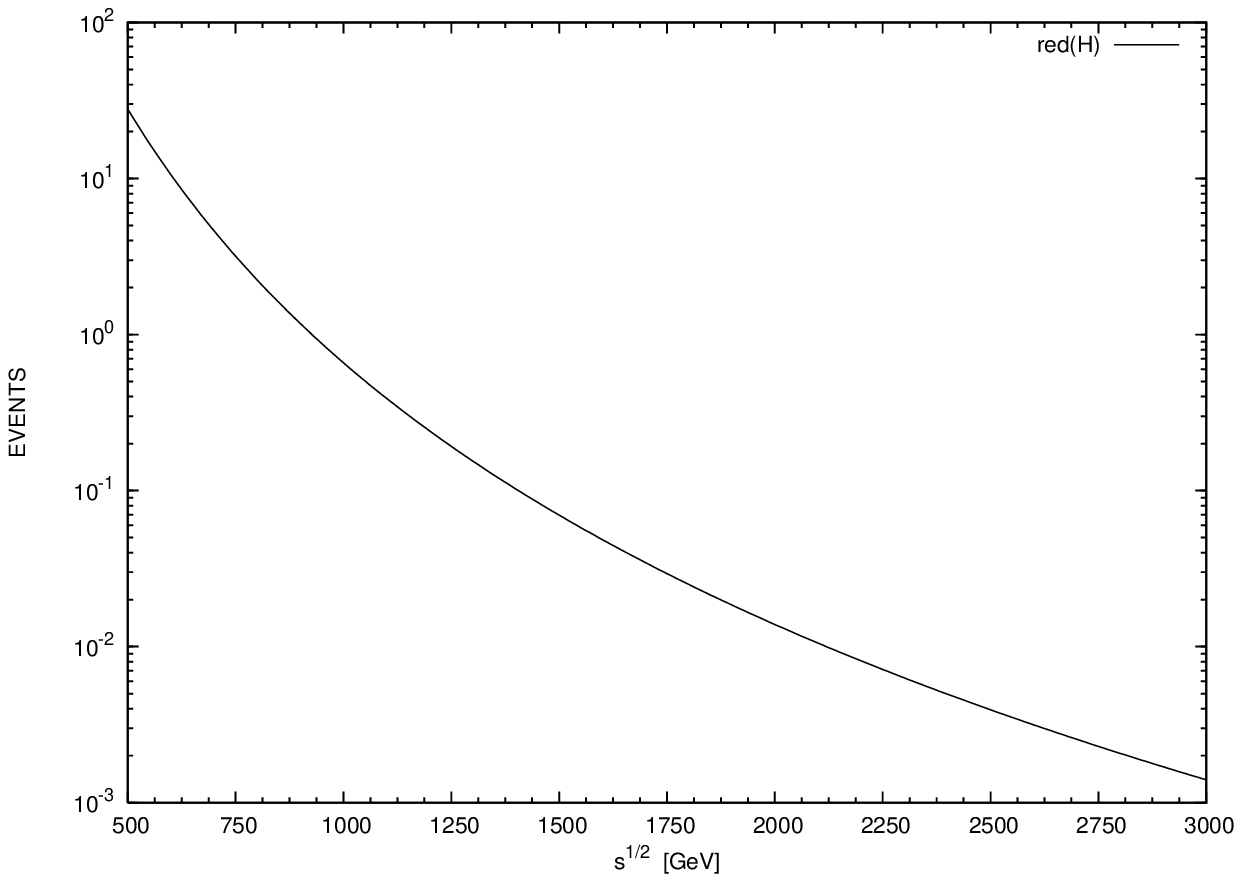}
\caption{\label{NE}The number of events  as a function of the Higgs mass for $\sqrt{s}=500$ GeV (left) and as a function of $\sqrt{s}$ for $m_H=115$ GeV (right)..}
\end{figure}

In Fig.~\ref{BTD} the contribution from diagrams displayed in Fig.~\ref{BT} is shown as a function of the Higgs mass and the center of mass energy of the collider, whereas in Fig.~\ref{HRD} the contribution arising from diagrams of Fig.~\ref{HR} are shown. From these figures, it can be appreciated that the contribution from diagrams of Fig.~\ref{BT} are suppressed by about 7 orders of magnitude compared with the contribution given by diagrams of Fig.~\ref{HR}. For $\sqrt{s}=500$ GeV, the respective contribution to the cross section can reach a value of $10^{-1}\, fb\,$ for $m_H\approx 140$ GeV, which shows clearly the marginal role played by diagrams of Fig.~\ref{BT}. However, this situation may be different in those processes involving heavier fermions, as the $\gamma \gamma \to tc$ reaction~\cite{WP}, with $t$ and $c$ the top and charm quarks, respectively. Finally, the number of events as a function of the Higgs mass and the center of mass energy of the collider is presented in Fig.~\ref{NE}. From this figure, it can be appreciated that up to $50$ events can be produced with a luminosity of  $\mathcal{L}_{e^+e^-}=500\, fb^{-1}$.

\section{Conclusions}
\label{co}In the minimal standard model the couplings of the Higgs boson to the remaining massive particles are thoroughly determined, which should be considered an outstanding feature of the model. However, there are well--motivated extensions of the model that require the presence of extended Higgs sectors. The extended theory is of course less predictive due to the proliferation of free parameters. The bonus is the appearance of interesting new physics effects such as lepton flavor violating transitions that are mediated by the physical scalars of the theory. In this paper, the potential of extended Yukawa sectors in generating  flavor violation both in the lepton and the quark sectors was explored in a model independent manner using the effective Lagrangian approach. A Yukawa sector that includes $SU_L(2)\times U_Y(1)$--invariant operators of up to dimension-six, which reproduces the main features of extended Yukawa sectors, was considered. The one--loop contribution of a flavor violating $Hf_if_j$ coupling, where $f_i$ and $f_j$ stand for charged leptons or quarks, to the on--shell $\gamma f_if_j$ and $\gamma \gamma f_if_j$ electromagnetic interactions was calculated. Exact formulae for the amplitudes associated with these couplings were presented and used to investigate electromagnetic lepton flavor violating transitions. The experimental limit on the branching ratio for the $\mu \to e\gamma$ decay was used to impose the bound $|\Omega_{\mu e}|<7\times 10^{-3}$ on the $H\mu e$ coupling. This bound allowed us to
derive the limit $Br(\mu \to e\gamma \gamma)<10^{-16}$, which is about  6 orders of magnitude more stringent than the corresponding experimental limit. The possibility of detecting signals of lepton flavor violation at the ILC through the reaction $e^+e^-\to \gamma \gamma \to \tau \mu$ was explored. It was found that with the projected luminosity of this machine up to half a hundred of $\tau^\pm \mu^\mp$ events could be produced. Although in some processes this number of events could be insufficient to detect the signal, it is important to have in mind that the $\tau^\pm \mu^\mp$ signal is a clean environment and is spectacular in the sense that the process is free of background, having thus the chance of being observed.

\acknowledgments{We acknowledge financial support from CONACYT and
SNI (M\' exico).}

\appendix*

\section{Form factors of the $\gamma \gamma f_if_j$ coupling}
In this appendix, we present the expressions for the form factors $F_{n _{R,L}}$. We present results only for the right--handed form factors $F_{n_{R}}$, as it can be passed to the left--handed ones via the interchanges $\omega_{k l}\omega_{m k}\leftrightarrow \omega^*_{k l}\omega^*_{m k}$ and $\omega^*_{k l}\omega_{m k}\leftrightarrow \omega_{k l}\omega^*_{m k}$. There are a total of 20 form factors of this class, but they are linked by the Bose symmetry, so one needs to list only 11 of them, the rest being obtained through the interchange $k_1\leftrightarrow k_2$. Also, the arguments of the diverse Passarino--Veltman coefficients are listed in Tables \ref{passarinoesc}, \ref{passarino1}, \ref{passarino2} y \ref{passarino3}. These form factors are given by:

\begin{align}
F_{1_R}=\,&4\,{\omega_{ i k} }\,m_W\,\left( {m_k }\,{\omega_{ k j}}\,
     \left( {D_{00}}(1) + {D_{00}}(2) \right)  -
    {m_j}\,{\omega^*_{k j}}\,\left( {D_{002}}(1) + {D_{002}}(2) \right)  \right),\nonumber\\
F_{2_R}=\,&\frac{{\omega_{ i k} }\,m_W^3}{2}\Bigg\{ 8\,\left( {m_k }\,{\omega_{ k j}}\,
          \left( {D_{22}}(1) + {D_{22}}(2) \right)  -
         {m_j}\,{\omega^*_{k j}}\,\left( {D_{222}}(1) + {D_{222}}(2) \right)  \right)  +
      \frac{1}{g(k_1,k_2)}\Big[{m_j}\left( {m_H^2} - {m_k^2 } \right) \,{\omega^*_{k j}}\nonumber\\
      &\times
          \left( {k_1} + {k_2}  \right)\cdot p_j \,{B_0}({i_1})  +
         4\,\left( {k_1}\cdot{k_2} + \left( {k_1} + {k_2}  \right)\cdot p_j \right) \,
          \left( {{m_j^2}} + 2\,\left({k_1}\cdot{k_2} +
            \left( {k_1} + {k_2}  \right)\cdot p_j\right) \right) \,
          \Big( \big( {m_k }\,{\omega_{ k j}}\,
                ( {C_0}(4)\nonumber\\
                &+ {C_1}(6) + {C_2}(6) )  +
               {m_j}\,{\omega^*_{k j}}\,
                \left( {C_0}(4) + 2\,{C_1}(6) + {C_{11}}(4) +
                  2\,\left( {C_{12}}(4) + {C_2}(6) \right)  + {C_{22}}(4) \right)  \big) \,
             {k_1}\cdot p_j\nonumber\\
             &+
            \big( {m_k }\,{\omega_{k j}}\,
                \left( {C_0}(3) + {C_1}(5) + {C_2}(5) \right)  +
               {m_j}\,{\omega ^*_{k j}}\,
                \left( {C_0}(3) + 2\,{C_1}(5) + {C_{11}}(3) +
                  2\,\left( {C_{12}}(3) + {C_2}(5) \right)  + {C_{22}}(3) \right)  \big) \nonumber\\
             &\times {k_2}\cdot p_j \Big)  +
         {B_0}({i_4})\,\left( {k_1} + {k_2} \right)\cdot p_j \,
          \Big( {m_j}\,\left( \left( {{m_j^2}} + {{m_k^2 }} -{{m_H^2}}  \right) \,
                {\omega^*_{k j}} + 2\,{m_j}\,{m_k }\,{\omega_{k j}} \right)  +
            2\,\left( {m_j}\,{\omega^*_{k j}} + 2\,{m_k }\,{\omega_{k j}} \right) \nonumber\\
             &\times\left( {k_1}\cdot{k_2} + ({k_1} +
               {k_2})\cdot p_j \right)  \Big) \Big] \Bigg\}            ,\nonumber\\
F_{3_R}=\,& -4\,{\omega^*_{k j}}\,{\omega_{ i k} }\,m_W^2\,
  \left( {D_{00}}(1) + {D_{001}}(1) + {D_{001}}(2) + {D_{003}}(1) \right)               ,\nonumber
\end{align}

\noindent where

\begin{equation}
g(k_1,k_2)=\,{k_1}\cdot p_j\, {k_2}\cdot p_j\,\left({k_1}\cdot{k_2} +
           ({k_1}+{k_2})\cdot p_j  \right) \,
         \left( {{m_j^2}} + 2\,\left({k_1}\cdot{k_2} +
           ({k_1}+{k_2})\cdot p_j\right) \right),\nonumber
\end{equation}

\begin{table}
\begin{center}
\begin{tabular}{@{}ccccccccccc@{}}
\hline $(j)$&1&2&3&4&5&6&7&8&9&10\\ \hline
$B_0(i_0)$&$0$&$m_H^2$&$m_H^2$&$$&$$&$$&$$&$$&$$&$$\\
$B_0(i_1)$&$0$&$m_H^2$&$m_k^2$&$$&$$&$$&$$&$$&$$&$$\\
$B_0(i_2)$&$0$&$m_k^2$&$m_k^2$&$$&$$&$$&$$&$$&$$&$$\\
$B_0(i_3)$&$m_j^2$&$m_H^2$&$m_k^2$&$$&$$&$$&$$&$$&$$&$$\\
$B_0(i_4)$&$m_j^2+h(k_1,k_2)$&$m_H^2$&$m_k^2$&$$&$$&$$&$$&$$&$$&$$\\
$B_0(1)$&$m_j^2+2\,k_1\cdot p_j$&$m_H^2$&$m_k^2$&$$&$$&$$&$$&$$&$$&$$\\
$B_0(2)$&$m_j^2+2\,k_2\cdot p_j$&$m_H^2$&$m_k^2$&$$&$$&$$&$$&$$&$$&$$\\
$C_0(i)$&$0$&$0$&$2\,k_1\cdot k_2$&$m_W^2$&$m_W^2$&$m_W^2$&$$&$$&$$&$$\\
$C_0(II)$&$0$&$0$&$2\,k_1\cdot k_2$&$m_t^2$&$m_t^2$&$m_t^2$&$$&$$&$$&$$\\
$C_0(1)$&$0$&$m_j^2$&$m_j^2+2\,k_1\cdot p_j$&$m_k^2$&$m_k^2$&$m_H^2$&$$&$$&$$&$$\\
$C_0(2)$&$0$&$m_j^2$&$m_j^2+2\,k_2\cdot p_j$&$m_k^2$&$m_k^2$&$m_H^2$&$$&$$&$$&$$\\
$C_0(3)$&$0$&$m_j^2+2\,k_1\cdot p_j$&$m_j^2+h(k_1,k_2)$&$m_k^2$&$m_k^2$&$m_H^2$&$$&$$&$$&$$\\
$C_0(4)$&$0$&$m_j^2+2\,k_2\cdot p_j$&$m_j^2+h(k_1,k_2)$&$m_k^2$&$m_k^2$&$m_H^2$&$$&$$&$$&$$\\
$D_0(1)$&$2\,k_1\cdot k_2$&$m_j^2+h(k_1,k_2)$&$m_j^2+2\,k_1\cdot p_j$&$0$&$m_j^2$&$0$&$m_k^2$&$m_k^2$&$m_H^2$&$m_k^2$\\
$D_0(2)$&$2\,k_1\cdot k_2$&$m_j^2+h(k_1,k_2)$&$m_j^2+2\,k_2\cdot p_j$&$0$&$m_j^2$&$0$&$m_k^2$&$m_k^2$&$m_H^2$&$m_k^2$\\
\hline
\end{tabular}
\caption{Arguments for the Passarino--Veltman scalar functions. The expression
has been  defined as: $h(k_1,k_2)=2\,(k_1\cdot k_2+(k_1+k_2)\cdot p_j)$.} \label{passarinoesc}
\end{center}
\end{table}

\begin{table}
\begin{center}
\begin{tabular}{@{}ccccccccccc@{}}
\hline $(j)$&1&2&3&4&5&6&7&8&9&10\\ \hline
$C_1(i_0)$&$m_j^2+h(k_1,k_2)$&$2\,k_1\cdot k_2$&$m_j^2$&$m_H^2$&$m_k^2$&$m_k^2$&$$&$$&$$&$$\\
$C_1(1)$&$m_j^2$&$0$&$m_j^2+2\,k_1\cdot p_j$&$m_H^2$&$m_k^2$&$m_k^2$&$$&$$&$$&$$\\
$C_1(2)$&$m_j^2$&$0$&$m_j^2+2\,k_2\cdot p_j$&$m_H^2$&$m_k^2$&$m_k^2$&$$&$$&$$&$$\\
$C_1(3)$&$m_j^2+2\,k_1\cdot p_j$&$0$&$m_j^2$&$m_H^2$&$m_k^2$&$m_k^2$&$$&$$&$$&$$\\
$C_1(4)$&$m_j^2+2\,k_2\cdot p_j$&$0$&$m_j^2$&$m_H^2$&$m_k^2$&$m_k^2$&$$&$$&$$&$$\\
$C_1(5)$&$m_j^2+2\,k_1\cdot p_j$&$0$&$m_j^2+h(k_1,k_2)$&$m_H^2$&$m_k^2$&$m_k^2$&$$&$$&$$&$$\\
$C_1(6)$&$m_j^2+2\,k_2\cdot p_j$&$0$&$m_j^2+h(k_1,k_2)$&$m_H^2$&$m_k^2$&$m_k^2$&$$&$$&$$&$$\\
$C_2(1)$&$m_j^2$&$0$&$m_j^2+2\,k_1\cdot p_j$&$m_H^2$&$m_k^2$&$m_k^2$&$$&$$&$$&$$\\
$C_2(2)$&$m_j^2$&$0$&$m_j^2+2\,k_2\cdot p_j$&$m_H^2$&$m_k^2$&$m_k^2$&$$&$$&$$&$$\\
$C_2(3)$&$m_j^2+2\,k_1\cdot p_j$&$0$&$m_j^2$&$m_H^2$&$m_k^2$&$m_k^2$&$$&$$&$$&$$\\
$C_2(4)$&$m_j^2+2\,k_2\cdot p_j$&$0$&$m_j^2$&$m_H^2$&$m_k^2$&$m_k^2$&$$&$$&$$&$$\\
$C_2(5)$&$m_j^2+2\,k_1\cdot p_j$&$0$&$m_j^2+h(k_1,k_2)$&$m_H^2$&$m_k^2$&$m_k^2$&$$&$$&$$&$$\\
$C_2(6)$&$m_j^2+2\,k_2\cdot p_j$&$0$&$m_j^2+h(k_1,k_2)$&$m_H^2$&$m_k^2$&$m_k^2$&$$&$$&$$&$$\\
$D_1(1)$&$2\,k_1\cdot k_2$&$m_j^2+h(k_1,k_2)$&$m_j^2+2\,k_1\cdot p_j$&$0$&$m_j^2$&$0$&$m_k^2$&$m_k^2$&$m_H^2$&$m_k^2$\\
$D_1(2)$&$2\,k_1\cdot k_2$&$m_j^2+h(k_1,k_2)$&$m_j^2+2\,k_2\cdot p_j$&$0$&$m_j^2$&$0$&$m_k^2$&$m_k^2$&$m_H^2$&$m_k^2$\\
$D_2(1)$&$2\,k_1\cdot k_2$&$m_j^2+h(k_1,k_2)$&$m_j^2+2\,k_1\cdot p_j$&$0$&$m_j^2$&$0$&$m_k^2$&$m_k^2$&$m_H^2$&$m_k^2$\\
$D_2(2)$&$2\,k_1\cdot k_2$&$m_j^2+h(k_1,k_2)$&$m_j^2+2\,k_2\cdot p_j$&$0$&$m_j^2$&$0$&$m_k^2$&$m_k^2$&$m_H^2$&$m_k^2$\\
\hline
\end{tabular}
\caption{ Arguments for the Passarino--Veltman tensorial coefficients with an index.} \label{passarino1}
\end{center}
\end{table}

\begin{table}
\begin{center}
\begin{tabular}{@{}ccccccccccc@{}}
\hline $(j)$&1&2&3&4&5&6&7&8&9&10\\ \hline
$C_{00}(1)$&$m_j^2+2\,k_1\cdot p_j$&$0$&$m_j^2$&$m_H^2$&$m_k^2$&$m_k^2$&$$&$$&$$&$$\\
$C_{00}(2)$&$m_j^2+2\,k_2\cdot p_j$&$0$&$m_j^2$&$m_H^2$&$m_k^2$&$m_k^2$&$$&$$&$$&$$\\
$C_{00}(3)$&$m_j^2+2\,k_1\cdot p_j$&$0$&$m_j^2+h(k_1,k_2)$&$m_H^2$&$m_k^2$&$m_k^2$&$$&$$&$$&$$\\
$C_{00}(4)$&$m_j^2+2\,k_2\cdot p_j$&$0$&$m_j^2+h(k_1,k_2)$&$m_H^2$&$m_k^2$&$m_k^2$&$$&$$&$$&$$\\
$C_{11}(1)$&$m_j^2+2\,k_1\cdot p_j$&$0$&$m_j^2$&$m_H^2$&$m_k^2$&$m_k^2$&$$&$$&$$&$$\\
$C_{11}(2)$&$m_j^2+2\,k_2\cdot p_j$&$0$&$m_j^2$&$m_H^2$&$m_k^2$&$m_k^2$&$$&$$&$$&$$\\
$C_{11}(3)$&$m_j^2+2\,k_1\cdot p_j$&$0$&$m_j^2+h(k_1,k_2)$&$m_H^2$&$m_k^2$&$m_k^2$&$$&$$&$$&$$\\
$C_{11}(4)$&$m_j^2+2\,k_2\cdot p_j$&$0$&$m_j^2+h(k_1,k_2)$&$m_H^2$&$m_k^2$&$m_k^2$&$$&$$&$$&$$\\
$C_{12}(1)$&$m_j^2+2\,k_1\cdot p_j$&$0$&$m_j^2$&$m_H^2$&$m_k^2$&$m_k^2$&$$&$$&$$&$$\\
$C_{12}(2)$&$m_j^2+2\,k_2\cdot p_j$&$0$&$m_j^2$&$m_H^2$&$m_k^2$&$m_k^2$&$$&$$&$$&$$\\
$C_{12}(3)$&$m_j^2+2\,k_1\cdot p_j$&$0$&$m_j^2+h(k_1,k_2)$&$m_H^2$&$m_k^2$&$m_k^2$&$$&$$&$$&$$\\
$C_{12}(4)$&$m_j^2+2\,k_2\cdot p_j$&$0$&$m_j^2+h(k_1,k_2)$&$m_H^2$&$m_k^2$&$m_k^2$&$$&$$&$$&$$\\
$C_{22}(1)$&$m_j^2+2\,k_1\cdot p_j$&$0$&$m_j^2$&$m_H^2$&$m_k^2$&$m_k^2$&$$&$$&$$&$$\\
$C_{22}(2)$&$m_j^2+2\,k_2\cdot p_j$&$0$&$m_j^2$&$m_H^2$&$m_k^2$&$m_k^2$&$$&$$&$$&$$\\
$C_{22}(3)$&$m_j^2+2\,k_1\cdot p_j$&$0$&$m_j^2+h(k_1,k_2)$&$m_H^2$&$m_k^2$&$m_k^2$&$$&$$&$$&$$\\
$C_{22}(4)$&$m_j^2+2\,k_2\cdot p_j$&$0$&$m_j^2+h(k_1,k_2)$&$m_H^2$&$m_k^2$&$m_k^2$&$$&$$&$$&$$\\
$D_{00}(1)$&$2\,k_1\cdot k_2$&$m_j^2+h(k_1,k_2)$&$m_j^2+2\,k_1\cdot p_j$&$0$&$m_j^2$&$0$&$m_k^2$&$m_k^2$&$m_H^2$&$m_k^2$\\
$D_{00}(2)$&$2\,k_1\cdot k_2$&$m_j^2+h(k_1,k_2)$&$m_j^2+2\,k_2\cdot p_j$&$0$&$m_j^2$&$0$&$m_k^2$&$m_k^2$&$m_H^2$&$m_k^2$\\
$D_{12}(1)$&$2\,k_1\cdot k_2$&$m_j^2+h(k_1,k_2)$&$m_j^2+2\,k_1\cdot p_j$&$0$&$m_j^2$&$0$&$m_k^2$&$m_k^2$&$m_H^2$&$m_k^2$\\
$D_{12}(2)$&$2\,k_1\cdot k_2$&$m_j^2+h(k_1,k_2)$&$m_j^2+2\,k_2\cdot p_j$&$0$&$m_j^2$&$0$&$m_k^2$&$m_k^2$&$m_H^2$&$m_k^2$\\
$D_{22}(1)$&$2\,k_1\cdot k_2$&$m_j^2+h(k_1,k_2)$&$m_j^2+2\,k_1\cdot p_j$&$0$&$m_j^2$&$0$&$m_k^2$&$m_k^2$&$m_H^2$&$m_k^2$\\
$D_{22}(2)$&$2\,k_1\cdot k_2$&$m_j^2+h(k_1,k_2)$&$m_j^2+2\,k_2\cdot p_j$&$0$&$m_j^2$&$0$&$m_k^2$&$m_k^2$&$m_H^2$&$m_k^2$\\
$D_{23}(1)$&$2\,k_1\cdot k_2$&$m_j^2+h(k_1,k_2)$&$m_j^2+2\,k_1\cdot p_j$&$0$&$m_j^2$&$0$&$m_k^2$&$m_k^2$&$m_H^2$&$m_k^2$\\
$D_{23}(2)$&$2\,k_1\cdot k_2$&$m_j^2+h(k_1,k_2)$&$m_j^2+2\,k_2\cdot p_j$&$0$&$m_j^2$&$0$&$m_k^2$&$m_k^2$&$m_H^2$&$m_k^2$\\
\hline
\end{tabular}
\caption{Arguments for the Passarino-Veltman tensorial coefficients with two indices.} \label{passarino2}
\end{center}
\end{table}

\begin{table}
\begin{center}
\begin{tabular}{@{}ccccccccccc@{}}
\hline $(j)$&1&2&3&4&5&6&7&8&9&10\\ \hline
$D_{001}(1)$&$2\,k_1\cdot k_2$&$m_j^2+h(k_1,k_2)$&$m_j^2+2\,k_1\cdot p_j$&$0$&$m_j^2$&$0$&$m_k^2$&$m_k^2$&$m_H^2$&$m_k^2$\\
$D_{001}(2)$&$2\,k_1\cdot k_2$&$m_j^2+h(k_1,k_2)$&$m_j^2+2\,k_2\cdot p_j$&$0$&$m_j^2$&$0$&$m_k^2$&$m_k^2$&$m_H^2$&$m_k^2$\\
$D_{002}(1)$&$2\,k_1\cdot k_2$&$m_j^2+h(k_1,k_2)$&$m_j^2+2\,k_1\cdot p_j$&$0$&$m_j^2$&$0$&$m_k^2$&$m_k^2$&$m_H^2$&$m_k^2$\\
$D_{002}(2)$&$2\,k_1\cdot k_2$&$m_j^2+h(k_1,k_2)$&$m_j^2+2\,k_2\cdot p_j$&$0$&$m_j^2$&$0$&$m_k^2$&$m_k^2$&$m_H^2$&$m_k^2$\\
$D_{003}(1)$&$2\,k_1\cdot k_2$&$m_j^2+h(k_1,k_2)$&$m_j^2+2\,k_1\cdot p_j$&$0$&$m_j^2$&$0$&$m_k^2$&$m_k^2$&$m_H^2$&$m_k^2$\\
$D_{003}(2)$&$2\,k_1\cdot k_2$&$m_j^2+h(k_1,k_2)$&$m_j^2+2\,k_2\cdot p_j$&$0$&$m_j^2$&$0$&$m_k^2$&$m_k^2$&$m_H^2$&$m_k^2$\\
$D_{122}(1)$&$2\,k_1\cdot k_2$&$m_j^2+h(k_1,k_2)$&$m_j^2+2\,k_1\cdot p_j$&$0$&$m_j^2$&$0$&$m_k^2$&$m_k^2$&$m_H^2$&$m_k^2$\\
$D_{122}(2)$&$2\,k_1\cdot k_2$&$m_j^2+h(k_1,k_2)$&$m_j^2+2\,k_2\cdot p_j$&$0$&$m_j^2$&$0$&$m_k^2$&$m_k^2$&$m_H^2$&$m_k^2$\\
$D_{222}(1)$&$2\,k_1\cdot k_2$&$m_j^2+h(k_1,k_2)$&$m_j^2+2\,k_1\cdot p_j$&$0$&$m_j^2$&$0$&$m_k^2$&$m_k^2$&$m_H^2$&$m_k^2$\\
$D_{222}(2)$&$2\,k_1\cdot k_2$&$m_j^2+h(k_1,k_2)$&$m_j^2+2\,k_2\cdot p_j$&$0$&$m_j^2$&$0$&$m_k^2$&$m_k^2$&$m_H^2$&$m_k^2$\\
$D_{223}(1)$&$2\,k_1\cdot k_2$&$m_j^2+h(k_1,k_2)$&$m_j^2+2\,k_1\cdot p_j$&$0$&$m_j^2$&$0$&$m_k^2$&$m_k^2$&$m_H^2$&$m_k^2$\\
$D_{223}(2)$&$2\,k_1\cdot k_2$&$m_j^2+h(k_1,k_2)$&$m_j^2+2\,k_2\cdot p_j$&$0$&$m_j^2$&$0$&$m_k^2$&$m_k^2$&$m_H^2$&$m_k^2$\\
\hline
\end{tabular}
\caption{Arguments for the Passarino--Veltman tensorial coefficients with three indices.} \label{passarino3}
\end{center}
\end{table}

\begin{align}
F_{5_R}=\,&           m_W^2\,\Bigg\{   2\,{\omega^*_{k j}}\,{\omega_{ i k} }\,
  \big( {C_0}(3) + 2\,{C_1}(5) + {C_{11}}(3) + 2\,{C_{12}}(3) +
    2\,{C_2}(5) + {C_{22}}(3) + 2\,\left( {D_{002}}(1) + {D_{002}}(2) \right)\nonumber\\
    &-
    2\,{D_{12}}(1)\,{k_1}\cdot{k_2} \big)   +   \frac{{\omega^*_{k j}}\,{\omega_{ i k} }\,}{{{k_1}\cdot p_j}}\Bigg[
     2\,{C_{00}}(3) + {{m_H^2}}\,
       \big( {C_0}(3) + {C_1}(5) + {C_2}(5) \big)  -
      {{m_k^2 }}\,\big( {C_0}(3) + {C_1}(5) + {C_2}(5) \big)\nonumber\\
      &+
      {{m_j^2}}\,\big( {C_0}(3) + 3\,{C_1}(5) + 2\,{C_{11}}(3) +
         4\,{C_{12}}(3) + 3\,{C_2}(5) + 2\,{C_{22}}(3) \big)  +
      2\,\big( {C_{12}}(3) + {C_2}(5) + {C_{22}}(3) \big) \,
       \big( {k_1}\cdot{k_2}\nonumber\\
       &+ {k_2}\cdot p_j \big) \Bigg]       +       \frac{{\omega^*_{k j}}\,{\omega_{ i k} }\,
    \left( {B_0}({i_4}) - 2\,{C_{00}}(4) \right) }{{k_2}\cdot p_j}     -     \frac{ {\omega^*_{k j}}\,{\omega_{ i k} }\,
      \left( {{m_H^2}}\,{B_0}({i_0}) -
        \left( {m_H^2} - {m_k^2 } \right) \,
         \left(  {B_0}({i_1}) -1  \right)  - {{m_k^2 }}\,{B_0}({i_2})
        \right)  }{2\,{k_1}\cdot p_j\,
    \left( {k_1}\cdot{k_2} + ({k_1} +
      {k_2})\cdot p_j \right) }\nonumber\\
      &-    \frac{{\omega^*_{k j}}\,{\omega_{ i k }}\,
       \left( {{m_H^2}} - {{m_k^2 }} + {{m_H^2}}\,{B_0}({i_0}) -
         {{m_k^2 }}\,{B_0}({i_2}) \right) }{\left( {{m_j^2}} - {{m_i^2}}
         \right) \,{k_2}\cdot p_j}  +
  \frac{1}{{m_j}\,\left( {m_j^2} - {m_i^2} \right) \,
     \left( {{m_j^2}} - {{m_i^2}} + 2\,{k_2}\cdot p_j \right) }\Bigg[\left( {m_H^2} - {m_k^2 } \right) \nonumber\\
      &\times\left( {m_j}\,{\omega^*_{k j}}\,{\omega_{ i k} } -
        {m_i}\,{\omega^*_{i k} }\,{\omega_{k j}} \right) \,{B_0}({i_1}) -
     \big( {m_j}\,{\omega^*_{k j}}\,
         \left( 2\,{m_i}\,{m_k }\,{\omega^*_{i k} } +
           \left(  {{m_j^2}} + {{m_k^2 }} -{{m_H^2}}  \right) \,{\omega_{ i k} } \right)
         + \big(
           {m_i}\,\left( {{m_j^2}} + {{m_k^2 }} \right) \,{\omega^*_{i k} }\nonumber\\
           &- {{m_H^2}}\,{m_i}\,{\omega^*_{i k} }  +
           2\,{{m_j^2}}\,{m_k }\,{\omega_{ i k }} \big) \,{\omega_{k j}} \big) \,
      {B_0}({i_3}) + 2\,{m_j}\,\left( {m_j^2} - {m_i^2} \right) \,
      \Big( {\omega_{ i k} }\,\big( {{m_H^2}}\,{\omega^*_{k j}}\,
            \left( {C_0}(2) + {C_1}(4) + {C_2}(4) \right)\nonumber\\
            &-
           {{m_k^2 }}\,{\omega^*_{k j}}\,
            \left( {C_0}(2) + {C_1}(4) + {C_2}(4) \right)  -
           {m_j}\,{m_k }\,{\omega_{k j}}\,
            \left( {C_0}(2) + {C_1}(4) + {C_2}(4) \right)  +
           {{m_j^2}}\,{\omega^*_{k j}}\,
            \big( {C_1}(2) + {C_{11}}(2) + 2\,{C_{12}}(2)\nonumber\\
            &+ {C_2}(2) +
              {C_{22}}(2) \big)  \big)  -
        {m_i}\,{\omega^*_{i k} }\,
         \Big( {m_k }\,{\omega^*_{k j}}\,
            \left( {C_0}(2) + {C_1}(4) + {C_2}(4) \right)  +
           {m_j}\,{\omega_{k j}}\,
            \big( {C_0}(2) + {C_1}(2) + {C_1}(4) + {C_{11}}(2)\nonumber\\
            &+
              2\,{C_{12}}(2) + {C_2}(2) + {C_2}(4) + {C_{22}}(2) \big)  \Big)
        \Big) \Bigg]               +      \frac{1}{2\,
    {k_2}\cdot p_j\,\left( {{m_j^2}} + 2\,{k_2}\cdot p_j \right) \,
    \left( {{m_j^2}} - {{m_i^2}} + 2\,{k_2}\cdot p_j \right) }\Bigg[ {m_j}\,\Big( {m_j}\,{\omega^*_{k j}}\nonumber\\
          &\times\left(
            \left(  {{m_H^2}} -  {{m_j^2}} - {{m_k^2 }}   \right) \,{\omega_{ i k} }  - 2\,{m_i}\,{m_k }\,{\omega^*_{i k} }  \right)
             + \left( {{m_H^2}}\,{m_i}\,{\omega^*_{i k} }   -
            {m_i}\,\left( {{m_j^2}} + {{m_k^2 }} \right) \,{\omega^*_{i k} } -
            2\,{{m_j^2}}\,{m_k }\,{\omega_{ i k} } \right) \,{\omega_{k j}} \Big) \,{B_0}(2)\nonumber\\
            &+ {m_j}\,\left( {m_H^2} - {m_k^2 } \right)  \,
     \left( {m_j}\,{\omega^*_{k j}}\,{\omega_{ i k} } -
       {m_i}\,{\omega^*_{i k} }\,{\omega_{k j}} \right) \,{B_0}({i_1}) -
    2\,\Big( 2\,{m_i}\,{m_k }\,{\omega^*_{i k} }\,{\omega^*_{k j}} +
       \left(  {{m_j^2}} + 2\,{{m_k^2 }} -2\,{{m_H^2}}  \right) \,{\omega^*_{k j}}\,
        {\omega_{ i k} }\nonumber\\
        &+ {m_j}\,
        \left( {m_i}\,{\omega^*_{i k} } + 2\,{m_k }\,{\omega_{ i k} } \right) \,
        {\omega_{k j}} \Big)\,{k_2}\cdot p_j \,{B_0}(2)\Bigg]       -        \frac{k_1\cdot k_2+(k_1+k_2)\cdot p_j}{2\,g(k_1,k_2)}\Bigg[ {\omega^*_{k j}}\,{\omega_{ i k} }\,
      \left( {k_1} + {k_2} \right)\cdot p_j \,
      \Big( \left( {m_H^2} - {m_k^2 } \right) \nonumber\\
         &\times{B_0}({i_1}) + {B_0}({i_4})\,
         \left(  {{m_j^2}} + {{m_k^2 }} -{{m_H^2}}  +
           2\,({k_1}\cdot{k_2} + ({k_1} +
           {k_2})\cdot p_j) \right)  \Big) \Bigg]                   \Bigg\}                   ,\nonumber\\
F_{7_R}=\,&-\frac{2\,{\omega^*_{k j}}\,{\omega_{ i k} }\,m_W^4}{{{k_2}\cdot p_j}}\Bigg\{
     {C_{12}}(4) + {C_2}(6) + {C_{22}}(4) +
      2\,\left( {D_{122}}(1) + {D_{122}}(2) + {D_{223}}(1) \right) \,
       {k_2}\cdot p_j  \Bigg\}                 ,\nonumber
\end{align}
\begin{align}
F_{9_R}=\,&     \frac{{\omega_{ i k} }\,m_W^3}{4}\Bigg\{ 8\,( {m_k }\,{\omega_{k j}}\,
          \left( {D_2}(1) + {D_2}(2) \right)  -
         {m_j}\,{\omega^*_{k j}}\,\left( {D_{22}}(1) + {D_{22}}(2) \right)  ) -
      \frac{1}{g(k_1,k_2)}\,\Big[{m_j}\,\left( {m_H^2} - {m_k^2 } \right) \,{\omega^*_{k j}}\nonumber\\
      &\times
          \left( {k_1} + {k_2} \right)\cdot p_j\,{B_0}({i_1})\,  +
         4\,\left( {k_1}\cdot{k_2} + ({k_1} +
            {k_2})\cdot p_j \right) \,
          \left( {{m_j^2}} + 2\,({k_1}\cdot{k_2} +
            ({k_1} + {k_2})\cdot p_j) \right) \,
          \Big( ( {m_k }\,{\omega_{k j}}\,{C_0}(4)\nonumber\\
          &+ {m_j}\,{\omega^*_{k j}}\,
                \left( {C_0}(4) + {C_1}(6) + {C_2}(6) \right)  ) \,
             {k_1}\cdot p_j +
            ( {m_k }\,{\omega_{k j}}\,
                ( {C_0}(3) + {C_1}(5) + {C_2}(5) )  +
               {m_j}\,{\omega^*_{k j}}\,
                ( {C_0}(3) + 2\,{C_1}(5)\nonumber\\
                &+ {C_{11}}(3) +
                  2\,\left( {C_{12}}(3) + {C_2}(5) \right)  + {C_{22}}(3) )  ) \,
             {k_2}\cdot p_j \Big)  +
         {B_0}({i_4})\,\left( {k_1} +
            {k_2} \right)\cdot p_j \,
          \Big( {m_j}\,( \left({{m_j^2}} + {{m_k^2 }} -{{m_H^2}}  \right) \,
                {\omega^*_{k j}}\nonumber\\
                &+ 2\,{m_j}\,{m_k }\,{\omega_{k j}} )  +
            2\,\left( {m_j}\,{\omega^*_{k j}} + 2\,{m_k }\,{\omega_{k j}} \right) \,
             \left( {k_1}\cdot{k_2} + ({k_1} +
               {k_2})\cdot p_j \right)  \Big) \Big]  \Bigg\}          ,\nonumber\\
F_{11_{R}}=\,&    \frac{{\omega^*_{k j}}\,{\omega_{ i k} }\,m_W^2}{{4\,{k_1}\cdot p_j}} \Bigg\{
     2\,\big(  {B_0}({i_4}) - {B_0}({i_2}) -
         {{m_H^2}}\,{C_0}(3) + {{m_k^2 }}\,{C_0}(3) -
         {{m_j^2}}\,\left( {C_0}(3) +
            2\,\left( {C_1}(5) + {C_2}(5) \right)  \right)  -
         2\,{C_2}(5)\,{k_1}\cdot{k_2} \big)\nonumber\\
         &-
      4\,\big( {C_0}(3) + {C_1}(5) + {C_1}({i_0}) + {C_2}(5) -
         2\,{D_{00}}(1) + 2\,{D_1}(1)\,{k_1}\cdot{k_2} \big) \,
       {k_1}\cdot p_j - 4\,{C_2}(5)\,{k_2}\cdot p_j \nonumber\\
       &+
      \frac{1}{{k_1}\cdot{k_2} + ({k_1} +
         {k_2})\cdot p_j}\Big[{{m_H^2}}\,{B_0}({i_0}) -
         \left( {m_H^2} - {m_k^2 } \right)  \,
          \left( {B_0}({i_1})  -1 \right)  - {{m_k^2 }}\,{B_0}({i_2})\Big]  \Bigg\}                   ,\nonumber\\
F_{13_{R}}=\,&    \frac{m_W^2}{{4}}\Bigg\{4\,{\omega^*_{k j}}\,{\omega_{ i k} }\,({C_1}({i_0}) - {C_2}(1) - 2\,{D_{00}}(1))  -
    \frac{\omega^*_{k j}\,{\omega_{ i k} }\,{B_0}({i_4})\,}{{k_1}\cdot p_j}\Bigg[ 1 - \frac{ {{m_k^2 }}-{{m_H^2}} }
          {{{m_j^2}} + 2\,({k_1}\cdot{k_2} +
            ({k_1} + {k_2})\cdot p_j)} \Bigg]\nonumber\\
            &+ \frac{1}{{k_1}\cdot p_j}\Bigg[\frac{1}{{m_j}}\Bigg[{m_j}\,{\omega^*_{k j}}\,{\omega_{ i k} }\,
           \left(  {B_0}(1) + \frac{2\,\left( {{m_H^2}} - {{m_k^2 }} \right) }
              {{{m_j^2}} - {{m_i^2}}} \right)  +
          \frac{2\,{{m_H^2}}\,{m_j}\,{\omega^*_{k j}}\,{\omega_{ i k} }\,
             {B_0}({i_0})}{{{m_j^2}} - {{m_i^2}}} +
          \frac{1}{{m_j^2} - {m_i^2}}\Bigg(\big( \left(   {m_k^2 } - {m_H^2}\right)  \nonumber\\
                &\times \left( {m_j}\,{\omega^*_{k j}}\,{\omega_{ i k} } -
                  {m_i}\,{\omega^*_{i k} }\,{\omega_{k j}} \right) \,
                {B_0}({i_1}) \big)  -
             2\,{m_j}\,\left( {{m_j^2}} - {{m_i^2}} + {{m_k^2 }} \right) \,
              {\omega^*_{k j}}\,{\omega_{ i k} }\,{B_0}({i_2}) +
             \big( {m_j}\,{\omega^*_{k j}}\,
                 \big( 2\,{m_i}\,{m_k }\,{\omega^*_{i k} }\nonumber\\
                 &+
                   \left( {{m_j^2}} + {{m_k^2 }} -{{m_H^2}}  \right) \,{\omega_{ i k} }
                   \big)  + \left(  {m_i}\,\left( {{m_j^2}} + {{m_k^2 }} \right) \,{\omega^*_{i k} } - {{m_H^2}}\,{m_i}\,{\omega^*_{i k} }  +
                   2\,{{m_j^2}}\,{m_k }\,{\omega_{ i k} } \right) \,{\omega_{k j}}
                \big) \,{B_0}({i_3})\Bigg) +
          2\,{m_j}\nonumber\\
          &\times \big( {m_i}\,{\omega^*_{i k} }\,
              \left( {m_k }\,{\omega^*_{k j}} + {m_j}\,{\omega_{k j}} \right)  +
             {\omega_{ i k} }\,\left(  {m_k }\,\left( {m_k }\,{\omega^*_{k j}} +
                   {m_j}\,{\omega_{k j}} \right)  - {{m_H^2}}\,{\omega^*_{k j}}  \right)  \big) \,{C_0}(1) +
          4\,{m_j}\,{\omega^*_{k j}}\,{\omega_{ i k} }\,{C_{00}}(1)\nonumber\\
          &+
          4\,{m_j}\,{\omega^*_{k j}}\,{\omega_{ i k} }\,{C_{00}}(3) -
          2\,{{m_j^2}}\,\left( {m_j}\,{\omega^*_{k j}}\,{\omega_{ i k} } -
             {m_i}\,{\omega^*_{i k} }\,{\omega_{k j}} \right) \,{C_1}(1) +
          2\,{m_j}\,{m_i}\,\left( {m_j}\,{\omega^*_{i k} }\,{\omega_{k j}}
             -\left( {m_i}\,{\omega^*_{k j}}\,{\omega_{ i k} } \right)   \right) \,{C_2}(1)\nonumber\\
             &+ \frac{1}{{m_i}\,\left( {{m_i^2}}  -{{m_j^2}}  -
               2\,{k_1}\cdot p_j \right)}\Bigg( {m_j}\,
                \Big( {m_i}\,{\omega^*_{k j}}\,
                   \left(
                     \left({{m_H^2}} -{{m_i^2}} - {{m_k^2 }}    \right) \,
                      {\omega_{ i k }}  -2\,{m_i}\,{m_k }\,{\omega^*_{i k} }  \right) -
                  {m_j}\,\big(
                     \left( {{m_i^2}} + {{m_k^2 }} \right) \,{\omega^*_{i k} }\nonumber\\
                     &- {{m_H^2}}\,{\omega^*_{i k }} +
                     2\,{m_i}\,{m_k }\,{\omega_{ i k} } \big) \,{\omega_{k j}}
                  \Big) \,{B_0}(1)   -
             \left( {m_j^2} - {m_i^2} \right) \, \left( {m_H^2} - {m_k^2 } \right) \,
              {\omega^*_{i k} }\,{\omega_{k j}}\,{B_0}({i_1}) +
             {m_i}\,\Big( 2\,{m_j}\,\left( {m_i^2} -{m_j^2}  \right)  \nonumber\\
             &\times {\omega^*_{k j}}\,{\omega_{ i k }}\,{B_0}({i_2}) +
                \Big( {m_j}\,{\omega^*_{k j}}\,
                    \left( 2\,{m_i}\,{m_k }\,{\omega^*_{i k} } +
                      \left( {{m_j^2}} + {{m_k^2 }} -{{m_H^2}}  \right) \,
                       {\omega_{ i k} } \right)  +
                   \big( {m_i}\,\left( {{m_j^2}} + {{m_k^2 }} \right) \,{\omega^*_{i k} } - {{m_H^2}}\,{m_i}\,{\omega^*_{i k} }\nonumber\\
                   &+
                      2\,{{m_j^2}}\,{m_k }\,{\omega_{ i k }} \big) \,{\omega_ {k j}}
                   \Big) \,{B_0}({i_3}) -
                2\,{m_j}\,\left( {m_j^2} - {m_i^2} \right)  \,
                 \Big( {\omega_{ i k} }\,
                    \big( {{m_H^2}}\,{\omega^*_{k j}}\,{C_0}(1) -
                      {{m_k^2 }}\,{\omega^*_{k j}}\,{C_0}(1) -
                      {m_j}\,{m_k }\,{\omega_{k j}}\,{C_0}(1)\nonumber\\
                      &-
                      2\,{\omega^*_{k j}}\,{C_{00}}(1) +
                      {{m_j^2}}\,{\omega^*_{k j}}\,{C_1}(1) \big)  +
                   {{m_i^2}}\,{\omega^*_{k j}}\,{\omega_{ i k} }\,{C_2}(1) -
                   {m_i}\,{\omega^*_{jk} }\,
                    \big( {m_k }\,{\omega^*_{k j}}\,{C_0}(1) +
                      {m_j}\,{\omega_{k j}}\,
                       \big( {C_0}(1) + {C_1}(1)\nonumber\\
                       &+ {C_2}(1) \big)  \big)  \Big)
                \Big) \Bigg) +
          \frac{{{m_j^2}}\,\left( {m_H^2} - {m_k^2 } \right)  \,{\omega^*_{jk} }\,{\omega_{k j}}\,
             \left( {B_0}(1) - {B_0}({i_1}) \right) }{{m_i}\,
             \left( {{m_j^2}} + 2\,{k_1}\cdot p_j \right) }\Bigg] +
       \frac{\left( {m_H^2} - {m_k^2 } \right) \,
          {\omega^*_{k j}}\,{\omega_{ jk} }\,{B_0}({i_1})}{{{m_j^2}} +
          2\,({k_1}\cdot{k_2} + ({k_1} +
          {k_2})\cdot p_j)}\Bigg]\Bigg\}                          ,\nonumber\\
F_{15_{R}}=\,&    -2\,{\omega^*_{k j}}\,{\omega_ {jk} }\,m_W^4\,\left( {D_{12}}(1) + {D_{23}}(1) \right)      ,\nonumber\\
F_{17_{R}}=\,&    -\frac{{\omega^*_{k j}}\,{\omega_{ jk} }\,m_W^4\,
      \left( {C_{12}}(3) + {C_2}(5) + {C_{22}}(3) -
        2\,{D_{12}}(1)\,{k_1}\cdot p_j \right) }{{k_1}\cdot p_j}                ,\nonumber\\
F_{19_{R}}=\,&        \frac{{\omega_{ jk }}\,m_W^3}{{8\, {k_1}\cdot p_j}}\Bigg\{ 4\,\big( {m_k }\,{\omega_{k j}}\,{C_0}(3) +
         {m_j}\,{\omega^*_{k j}}\,
          \left( {C_0}(3) + {C_1}(5) + {C_2}(5) \right)  \big)  +
      8\,\left( {m_k }\,{\omega_{k j}}\,{D_0}(1) -
         {m_j}\,{\omega^*_{k j}}\,{D_2}(1) \right) \,{k_1}\cdot p_j\nonumber\\
         &+
      \frac{k_1\cdot p_j\, k_2\cdot p_j}{g(k_1,k_2)}\Bigg[{m_j}\,\left( {m_H^2} - {m_k^2 } \right) \,{\omega^*_{k j}}\,{B_0}({i_1}) +
         {B_0}({i_4})\,\big( {m_j}\,
             \left( \left( {{m_j^2}} + {{m_k^2 }} -{{m_H^2}}\right) \,
                {\omega^*_{k j}} + 2\,{m_j}\,{m_k }\,{\omega_{k j}} \right)\nonumber\\
                &+
            2\,\left( {m_j}\,{\omega^*_{k j}} + 2\,{m_k }\,{\omega_{k j}} \right) \,
             \left( {k_1}\cdot{k_2} + ({k_1} +
               {k_2})\cdot p_j \right)  \big) \Bigg] \Bigg\}      ,\nonumber\\
F_{21}=\,&\frac{8\,m_W^2}{2\,k_1\cdot k_2}\left(3 + \frac{2\,k_1\cdot k_2}{2\,m_W^2} +
6\,m_W^2\left(1 - \frac{2\,k_1\cdot k_2}{2\,m_W^2}\right)\,C_0(i)\right)
-Q_t^2\,{N_{c_{t}}}\,\frac{8\,m_t^2}{2\,k_1\cdot k_2}
\left(2 + (4\,m_t^2 - 2\,k_1\cdot k_2)\,C_0(II)\right).
\end{align}

\end{document}